\pdfoutput=1

\documentclass[12pt]{article}

\usepackage{slashed}
\usepackage{color,verbatim}
\usepackage{latexsym}
\usepackage{epsfig}
\usepackage{amsmath,accents}
\usepackage[utf8]{inputenc}

\usepackage{cite}
\usepackage{soul}

\usepackage{amssymb}
\usepackage{graphicx}
\usepackage{bm}
   
\makeatletter

\usepackage{enumerate}

\@addtoreset{equation}{section}
\makeatother

\newcommand{\be}{\begin{equation}}
\newcommand{\ee}{\end{equation}}
\newcommand{\bea}{\begin{eqnarray}}
\newcommand{\eea}{\end{eqnarray}}

\newcommand{\Li}{\operatorname{Li}}

\begin{document}

\thispagestyle{empty}

\begin{center}

\begin{center}

\vspace{.5cm}

{\Large\bf Natural supersymmetry from extra dimensions}
\end{center}

\vspace{1.cm}

\textbf{
A. Delgado$^{\,a}$, M. Garcia-Pepin$^{\,b}$, G. Nardini$^{\,c}$, M. Quir\'os$^{\,b\,,d}$
}\\

\vspace{1.cm}
${}^a\!\!$ {\em {Department of Physics, University of Notre Dame\\Notre Dame, Indiana 46556, USA}}

\vspace{.1cm}
${}^b\!\!$ {\em {Institut de F\'{\i}sica d'Altes Energies (IFAE),\\ The Barcelona Institute of  Science and Technology (BIST),\\ Campus UAB, 08193 Bellaterra (Barcelona) Spain}}

\vspace{.1cm}
${}^c\!\!$ {\em {Albert Einstein Center (AEC), Institute for Theoretical Physics (ITP),\\ University of Bern, Sidlerstrasse 5, CH-3012 Bern, Switzerland}}

\vspace{.1cm}
${}^d\!\!$ {\em {Instituci\'o Catalana de Recerca i Estudis  
Avan\c{c}ats (ICREA), \\ Campus UAB, 08193 Bellaterra (Barcelona) Spain}}


\end{center}

\vspace{0.8cm}

\centerline{\bf Abstract}
\vspace{2 mm}
\begin{quote}\small
  We show that natural supersymmetry can be embedded in a
  five-dimensional theory with supersymmetry breaking \`a la
  Scherk-Schwarz (SS).  There is no `gluino-sucks' problem for stops
  localized in the four-dimensional brane and gluinos propagating in
  the full five-dimensional bulk, and sub-TeV stops are easily
  accommodated.  The $\mu/B_\mu$ problem is absent as well; the SS
  breaking generates a Higgsino Dirac mass and no bilinear Higgs mass
  parameter in the superpotential is required. Moreover, for
  nonmaximal SS twists leading to $\tan\beta\simeq 1$, the Higgs
  spectrum is naturally split, in agreement with LHC data. The 125 GeV
  Higgs mass and radiative electroweak symmetry breaking can be
  accommodated by minimally extending the Higgs sector with $Y=0$
  $SU(2)_L$ triplets.
  \end{quote}

\vfill

\newpage

\tableofcontents

\newpage

\section{Introduction}
\label{sec:intro}

In the past decades the hierarchy problem of the Standard Model (SM)
has guided most of the particle physics community in the search for a
UV completion able to describe nature up to the Planck (or GUT) cutoff
scale. In this task, supersymmetry and compositeness have been, and
still are, the most promising lighthouses to follow. Their most
appealing feature is that their Higgs sectors are insensitive to the
Planck mass cutoff, and are only sensitive to the scale of new physics
which should, therefore, be close to the electroweak scale in order to
avoid an (unnatural) {\it little} hierarchy problem. Despite this
expectation, there is no sign of new physics in the LHC data.  The
situation is thus threatening: if the tendency in the data does not
change, we might lose our trust in the naturalness criterion (and the
subsequent loss of confidence on any deduction based on dimensional
arguments), which would make our future way up to the Planck scale
very hard. In order to avoid this threat, it is
crucial to understand whether, and in case why, naturalness is hiding
in the present LHC data.

In supersymmetry, several experimental observations seem to invoke a
tuning in the electroweak sector. Indeed, if one does not rely on the
low-energy corners of the parameter space still compatible with the
experimental searches, a large gap between the soft-supersymmetry
breaking and electroweak scales is required~\cite{Aad:2016tuk,Khachatryan:2016uwr}. The tension
between data and naturalness is however reduced, and may be avoided, if
there is a symmetry imposing some cancellations at both tree level and
(at least) one loop.

The naturalness problem is
manifest in the minimal supersymmetric extension of the SM (MSSM).  In the MSSM the
squared-mass term of the lighter CP-even (and SM-like) eigenstate $h$ has the magnitude of the lighter eigenvalue in the matrix
\be
\mathcal M_{H_1,H_2}^2=\left(
\begin{array}{cc}
  m^2_1
  & m^2_3
   \\
  m^2_3
   & m_2^2
\end{array}
\right)~,
\label{eq:MSSMmass}
\ee
where $m_i^2=(m_i^2)^0+\Delta m_i^2$ contains radiative corrections
$\Delta m_i^2$ to the desired order. The lightest eigenvalue of
$\mathcal M_{H_1,H_2}^2$ thus needs to be $\mathcal O(m_Z^2)$ and
negative to have agreement with the experimentally observed
electroweak symmetry breaking (EWSB) pattern. The other Higgses,
instead, with a squared mass of the order of the larger eigenvalue of
$\mathcal M_{H_1,H_2}^2$, have to be hierarchically larger to avoid
any tension with the extra-Higgs
searches~\cite{Aad:2015kna,Khachatryan:2015tha}.  Two parameter
regions seem promising for fulfilling these features:
\begin{itemize}
\item
The so-called focus point
solution~\cite{Feng:1999zg,Feng:2011aa,Delgado:2014vha,Delgado:2014kqa,Masina:2015ixa},
based on the fact that, for $m_1^2\gg m_3^2$, or equivalently
$\tan\beta\gg 1$, $\mathcal M_{H_1,H_2}^2$ is essentially diagonal. In
this case no tree-level tuning is required if $m_2^2\sim
\mathcal O(-m_Z^2)$.
\item
The parameter region $m_1^2\simeq m_2^2\simeq m_3^2\gg \mathcal
O(m_Z^2)$ (equivalent to $\tan\beta\simeq 1$) which, if justified by a
symmetry, naturally leads to a vanishing eigenvalue in $\mathcal
M_{H_1,H_2}^2$.
\end{itemize}

However, several issues jeopardize the naturalness of these two
options.  In particular the supersymmetric parameter
$\mu$~\footnote{This parameter provides a supersymmetric contribution
  to the tree-level Higgs masses $(m_{1,2}^2)^0=m_{H_{1,2}}^2+\mu^2$.}
cannot be below the electroweak scale because of the lightest chargino
mass bound, $m_{\widetilde \chi^\pm}\gtrsim
105\,$GeV~\cite{Heister:2002mn}. Moreover, even if $\mu$ is above this
bound, it can be in tension with the general electroweakino searches,
depending on the gaugino mass
spectrum~\cite{Aad:2014yka,Khachatryan:2014qwa}. Finally if all these
constraints are circumvented, still, explaining theoretically why the
electroweak scale appears naturally in the superpotential is
challenging.

The radiative corrections $\Delta m_i^2$ should also lift
concerns. They should be small in order not to introduce a tuning at
one loop. In this sense, the charged sleptons and bottom squarks that
must be heavy to fulfill the flavor constraints~\cite{Isidori:2010kg},
are innocuous when $\tan\beta$ is not huge. Stop contributions are
instead dangerous. Thus, naturalness requires light stops, which are
in agreement with top squark searches and 125 GeV Higgs mass
constraints only in the presence of heavy gluinos and sizable stop
mixing~\cite{ATLAS-CONF-2015-067,CMS-PAS-SUS-15-003,
  Aaboud:2016lwz}. Unfortunately, the latter also generate large
radiative corrections that need to be tuned, while the former tend to
be inconsistent with light stops in top-down approaches. In fact,
heavy gluinos pull the stop soft masses above the TeV scale along the
running from the scale at which they are generated (if this scale is
large enough), to the electroweak scale~\cite{Martin:1993zk}.

In view of the above issues, an appropriate strategy to resurrect naturalness in
the present LHC data may consist in looking for UV embeddings where:
\begin{enumerate}[{\it (i)}]
\item
  The tree-level Higgs mass is higher than in the (``vanilla") MSSM in
  such a way that rather light stops with negligible mixing are viable.
  \item Gluinos are heavy, but the scale at which the soft stop masses
    are generated, and below which the renormalization-group
    (RG) evolution applies, is rather low (i.e.~stop masses remain
    small while running down to the electroweak scale).
\item
  In the superpotential no mass term is required, so
  that the Dirac mass of the Higgsinos does \emph{not} have a
  superpotential origin.
\end{enumerate}
  
Remarkably, the five-dimensional (5D) $N=1$ supersymmetry
embeddings of Refs.~\cite{Antoniadis:1990ew, Pomarol:1998sd} can fulfill these
requirements~\cite{Antoniadis:1998sd,Delgado:1998qr}. When the fifth
dimension is compactified on the circle orbifold $S_1/\mathbb Z_2$ of
radius $R$, the $N=1$ chiral superfields propagating in the bulk
receive either soft supersymmetry-breaking scalar masses and/or Dirac
fermionic masses, depending on some global charge assignments
technically called Scherk-Schwarz (SS)
twists~\cite{Scherk:1979zr,Quiros:2003gg}. It is then possible to
arrange the 5D Higgs sector in bulk chiral superfields in such a way
that, below the compactification scale, the four-dimensional (4D) effective theory is
equivalent to the MSSM with either $\tan\beta=1$ [with
  $m_1^2=m_2^2=m_3^2\sim\mathcal O(1/R^2)$ exact at tree level] for
nonmaximal
twists~\cite{Pomarol:1998sd,Antoniadis:1998sd,Delgado:1998qr} or
$\tan\beta=\infty$ for maximal
twists~\cite{Dimopoulos:2014aua,Garcia:2015sfa}.  Crucially, no
contribution mimicking a superpotential squared mass $\mu^2$
arises, although the Higgsinos do receive an $\mathcal O(1/R)$ Dirac
mass, as required by condition {\it (iii)}~\footnote{Notice that also
  the chiral superfields associated to the first and second generation
  of squarks and sleptons will benefit of the same supersymmetry breaking
  if these superfields propagate in the bulk. There exist charge
  assignments for which their fermions are massless (prior to EWSB)
  but their superpartners are at the $\mathcal O(1/R)$ scale. This
  makes the SS mechanism naturally compatible with the flavor
  constraints~\cite{Isidori:2010kg}.}.

This supersymmetry breaking mechanism, dubbed the SS
mechanism~\cite{Scherk:1979zr}, also works on vector superfields. It
naturally leads to a spectrum where all gauge bosons are massless
(prior to EWSB) and all gauginos have $\mathcal O(1/R)$ Majorana
masses. On the other hand, at tree level, it does not induce any
supersymmetry breaking for superfields localized at a brane of the
orbifold. Therefore, by assuming a localized third generation of
squarks, the stop soft squared masses are generated with a suppression
of a one-loop factor~\footnote{The stop mixing is also suppressed by a
  loop factor and far away from the maximal mixing value as discussed
  in Sec.~\ref{MSSM}.}. Moreover, since the logarithmic ratio
between the electroweak scale and the compactification scale (at which
the SS mechanism induces supersymmetry breaking) is small, the stop
masses are not drastically modified by their RG evolution and remain
$\mathcal O(0.1/R)$, as required by condition {\it (ii)}.
Finally, also the requirement \textit{(i)} can be fulfilled. In
(maximally twisted) SS scenarios leading to $\tan\beta=\infty$, 5D
nonminimal supersymmetric extensions with either one singlet on the
brane, or two pairs of vectorlike fermions on the brane, or an extra
$U(1)^\prime$ vector superfield in the bulk, boost the tree-level mass
of the SM-like
Higgs~\cite{Dimopoulos:2014aua,Garcia:2015sfa}~\footnote{At a
  quantitative level, the singlet case might be problematic as its
  $F$-term contribution to the Higgs quartic coupling is suppressed at
  large $\tan\beta$.}.

The 5D SS scenarios then contain all the ingredients guaranteeing the
SM-like Higgs squared-mass term to be $\mathcal O(m_Z^2)$, provided by gauge interactions 
without an unnatural tuning. The last obstacle is the sign of this term. In fact,
in the above SS scenarios solving the issue \textit{(i)}, EWSB
(namely with a negative sign in the Higgs squared mass term) can be achieved
only by means of higher-dimensional operators whose magnitude and
origin are hard to identify. It is then worth proving that there exist
SS scenarios where these operators are not necessary for a successful
EWSB, and where, at the same time, the requirements $(i)$, $(ii)$ and $(iii)$ are
fulfilled. We achieve this result by focusing on minimal extensions of
the chiral superfield sector (for a study where the EWSB is obtained
in nonminimal chiral extensions violating condition $(iii)$, see
Ref.~\cite{Cohen:2015gaa}).  We also restrict ourselves to the orbifold
charge assignments corresponding to the $\tan\beta =1$ case, for which
the $F$ terms contributions to the Higgs tree-level mass are enhanced.

Our proof of principle is developed in several steps. In
Sec.~\ref{MSSM} we review how to embed the MSSM in a 5D SS scenario
and why the 125 GeV Higgs mass and the EWSB are problematic. Since the
former problem should be trivially avoidable in a MSSM 5D scenario
supplemented by a singlet, we consider in Sec.~\ref{singlets} the
case where there is an extra singlet chiral superfield localized at
the brane. As expected, the Higgs mass bound can be accommodated, but
the radiative corrections to the SM-like squared-mass term are not
sufficient to trigger the EWSB, analogously to the $\tan\beta=\infty$
case~\cite{Dimopoulos:2014aua,Garcia:2015sfa}. In addition, and
surprisingly previously unnoticed, as the singlet is not protected by any symmetry of the theory,  it develops a large tadpole (prior to the EWSB) inducing a $\mathcal O(1/R)$ vacuum
expectation value (VEV) to the singlet. This jeopardizes the treatment
of the Higgs Kaluza-Klein (KK) towers and the possibility of achieving
Higgs-singlet mixings compatible with the LHC
constraints~\cite{Khachatryan:2016vau}. This VEV could of course be
suppressed by introducing a (huge) singlet mass term in the
superpotential~\footnote{This term was considered in
  e.g.~Ref.~\cite{Cohen:2015gaa} to suppress the effect of the tadpole
  estimated to be of the order of the electroweak scale.}. Since this
possibility would violate the criterion \textit{(iii)}, and moving the
singlet to the bulk should not circumvent the problem, in
Sec.~\ref{sec:triplets} we pursue the analysis with the $Y=0$
$SU(2)_L$-triplet extension of the MSSM, in which case the gauge symmetry
itself forbids the large tadpole.  This case, with the 5D $N=1$
triplet superfield being in the bulk triplet charginos would be too
light if the superfield were localized on the brane turns out to be
the example of SS scenario that satisfies all the requests of our
proof. Finally, in Sec.~\ref{sec:low} we discuss some further
phenomenological bounds and the need for localizing on the brane the
third family of the leptonic superfield to overcome the dark matter bound,
and in Sec.~\ref{sec:concl} we present our conclusions.

 \section{5D MSSM}
 \label{MSSM}
 We embed the MSSM in a 5D space-time setup where the extra dimension
 is the orbifold $S^1/\mathbb Z_2$ with two 4D
 branes at the fixed points $y=0$ and $y=\pi R$ ($R$ is the radius of
 the circle $S^1$).  The gauge and Higgs sectors, as well as the first
 and second generations of matter (and the right-handed
 stau~\footnote{We are considering $\widetilde\tau_R$ propagating in
   the bulk in order to avoid bounds on heavy stable charged
   particles~\cite{CMS-PAS-EXO-16-036}.
 }),
 propagate in the bulk while the rest of the third generation matter
 is localized at the $y=0$ brane. The boundary conditions of the bulk
 fields are twisted by introducing SS parameters associated with the
 available global symmetries we are allowed to break. In this section we present a
 summary of the formalism and results in the MSSM (nonminimal
 extensions are considered in Secs.~\ref{singlets} and
 \ref{sec:triplets}). The original calculations were performed mainly
 in Refs.~\cite{Pomarol:1998sd,Antoniadis:1998sd,Delgado:1998qr} to
 which we will refer for more details. To simplify the notation,
 hereby, unless explicitly stated, we use units where $R\equiv 1$.

In 5D supersymmetry the Higgs doublets in the bulk belong to the $N=2$
hypermultiplets $\mathbb H_a=(H_a,H_a^c,\Psi_a,F_a,F_a^c)$ (with
$a=1,2$), where $H_a$ and $H_a^c$ are complex $SU(2)_L$ doublets with
hypercharge $1/2$ and $\Psi_a=(\psi_a,\bar\psi_a^c)^T\equiv
(\psi_{aL},\psi_{aR})^T$ are $SU(2)_L$-doublet Dirac spinors with
$\psi_a$ ($\bar\psi_a$) and $\psi_a^c$ ($\bar\psi_a^c$) being
undotted (dotted) Weyl spinors. The two hypermultiplets $\mathbb H_a$
have the same quantum numbers and can then be arranged to form a
doublet of a global symmetry, $SU(2)_H$, acting on the index $a$.  The
doublet of $N=2$ hypermultiplets can also be split into $\mathbb Z_2$
even and odd $N=1$ chiral multiplets according to the $\mathbb Z_2$
parity assignment
\be
\mathbb Z_2=\left.\sigma_3\right|_{SU(2)_H}\otimes \gamma_5
\label{orbifold}
\ee
where $\sigma_3$ acts on the $SU(2)_H$ indices and $\gamma_5$ over Dirac indices.
For $\mathbb H_1$ and $\mathbb H_2$, we take $(H_2,\psi_{2L},F_2)$ and $(H_1,\psi_{1R},F_1)$ to be even and $(H_2^c,\psi_{2R},F_2^c)$ and $(H_1^c,\psi_{1L},F_1^c)$ to be odd. 

The gauge sector in the bulk is instead described by $N=2$ vector
supermultiplets. For a $SU(N)$ gauge group each of the
supermultiplets is given by $\mathbb V=(V_M, \lambda_L^i,\Upsilon)$,
which contains the vector bosons $V_M$ (with $M=1,\dots,5$), the real
scalar $\Upsilon$ and the two bispinors $\lambda^i_L$ (with
$i=1,2$). All these fields are in the adjoint representation of
$SU(N)$. As customary we assume $V_\mu$ and $\lambda_L^1$
($V_5$,$\Upsilon$ and $\lambda_L^2$) to be even (odd) with respect to
the $\mathbb Z_2$ symmetry.

The SS twists $(q_R,q_H)$ associated with the global symmetries  $SU(2)_R\times SU(2)_H$ impose the relation
%
%
%
 \be
 \left[\begin{array}{cc} H_1(x,y)& H_1^c(x,y)\\ H_2^c(x,y) & H_2(x,y) \end{array}\right]=e^{i q_H\sigma_2 y}
 \sum_{n=0}^\infty  \sqrt{\frac{2}{\pi}} \left[\begin{array}{cc} \cos ny\, H_1^{(n)}(x)& \sin ny\, H_1^{c(n)}(x)\\ \sin ny\, H_2^{c(n)}(x) & \cos ny\, H_2^{(n)}(x) \end{array}\right]
 e^{-i q_R\sigma_2 y}~,
 \label{doublets}
 \ee
where $H_{1,2}^{(n)}(x)$ (with $n\geq 0$) and $H_{1,2}^{c(n)}(x)$ (with $n\geq 1$) are the KK modes of the corresponding doublets (their $x$ dependence is omitted hereafter) and  have mass dimension equal to one. The $\sqrt{2/\pi}$ factor comes from the normalization of the nonzero modes in the interval $[0,\pi]$. The zero modes $H_a^{(0)}$ are then \textit{not} canonically normalized as they are missing a prefactor $1/\sqrt{2}$. The mass doublet eigenstates $h^{(n)}$ and $H^{(n)}$, with masses $q_R-q_H+n$ and $q_R+q_H+n$ respectively (with $n$ from $-\infty$ to $+\infty$), are computed in Ref.~\cite{Pomarol:1998sd}. They are given by
\begin{align}
H_1^{(n)}=&\left(h^{(n)}+h^{(-n)}+H^{(n)}+H^{(-n)}   \right)/2~,\nonumber\\
H_2^{(n)}=&\left(h^{(n)}+h^{(-n)}-H^{(n)}-H^{(-n)}   \right)/2~,\nonumber\\
H_1^{c(n)}=&\left(h^{(-n)}-h^{(n)}+H^{(-n)}-H^{(n)}   \right)/2~,\nonumber\\
H_2^{c(n)}=&\left(h^{(n)}-h^{(-n)}+H^{(-n)}-H^{(n)}   \right)/2~,
\label{nonceroH}
\end{align}
for $n\geq 1$, and by
\begin{align}
H_1^{(0)}=&\left( h^{(0)}+H^{(0)}  \right)/2~,\nonumber\\
H_2^{(0)}=&\left( h^{(0)}-H^{(0)}  \right)/2~,
\label{ceroH}
\end{align}
for $n=0$.  Notice that although $H_a^{(0)}$ are noncanonically
normalized, the zero modes $h^{(0)}$ and $H^{(0)}$ are canonically
normalized, which has enforced the introduction of an extra factor of $1/\sqrt{2}$ in
Eq.~(\ref{ceroH}). In this way, even though the zero modes are
differently normalized than the nonzero ones, it is straightforward
to reconstruct full KK towers (with $n$ from $-\infty$ to $+\infty$) of
fields when coupled to the brane.
As for the Higgsino components in $\mathbb H_a$, the mass
eigenstates are (for $n>0$)
\begin{align}
\widetilde H^{(-n)}= &\frac{1}{\sqrt{2}}\left( \psi_2^{(n)}-\gamma_5\psi_1^{(n)} \right),\ & \text{with mass }(q_H-n)\, ,\nonumber\\
\widetilde H^{(+n)}= &\frac{1}{\sqrt{2}}\left(\gamma_5 \psi_2^{(n)}+\psi_1^{(n)} \right),\  &\text{with mass }(q_H+n)\, ,\nonumber\\ 
\widetilde H^{(0)}= & \left(\psi_{2L}^{(0)}, \psi_{1R}^{(0)}\right)^T \, , &   \text{with mass }(q_H)\, . \quad~~\,
\label{eq:Hinos}
\end{align}

The $SU(2)_R$ twist also acts on all bulk gauginos, which are embedded
in $\mathbb V_j$. The KK tower of these fields have Majorana masses
$n+q_R$. Moreover, any field in the bulk coupled to these charginos is
sensitive to the twist $q_R$. All bulk matter fields have, in fact, a KK
tower with tree-level masses $n+q_R$ for scalars and $n$ for fermions.
Bulk fields that are $SU(2)_R$ singlets (e.g.~the gauge vector bosons
and scalars of $\mathbb V_j$), or fields in the brane, are instead
insensitive to $q_R$ and their spectrum is not affected by the SS
mechanism~\footnote{For our twist assignments, each bosonic component
  $V^j_\mu,V^j_5,\Upsilon^j$ of $\mathbb V^j$ exhibits a KK spectrum
  with masses $n$, and both $V_5$ and $\Upsilon$ have vanishing zero
  modes.}.

The scenario with charges $q_R=q_H\equiv \omega$ is particularly
interesting. The doublet $h^{(0)}$ is massless while the doublet
$H^{(0)}$ has mass $2\omega$. The corresponding KK modes, $h^{(n)}$
and $H^{(n)}$, have masses $n$ and $2\omega+n$,
respectively. Higgsinos and charginos have masses $\omega+n$. The first
and second generation sfermions and right-handed staus, which we assume
in the bulk, also have mass eigenstates $\omega+n$, while their
supersymmetric partners have masses $n$. In the rest of the paper we
focus on this scenario and some minimal extensions of it.

The main features of this scenario are the following:
\begin{itemize} 
   \item
     The Higgsino zero mode has a Dirac mass equal to $\omega/R$, by
     which there is no need to introduce a superpotential $\mu$-like
     term as in the MSSM. \textit{The $\mu$-problem is thus naturally
       solved by this formalism.}

   \item
     At tree level the theory predicts a 4D massless Higgs doublet
     with a flat potential~\cite{Pomarol:1998sd}. The rest of the
     Higgs sector is heavy. In the MSSM language this amounts to
     equations of electroweak minimum with $\tan\beta=1$ and invariant
     under the global scale change $\omega/R\to
     \lambda\ \omega/R$. Such an invariance reminds some of the
     properties of the focus point
     solution~\cite{Feng:1999zg,Feng:2011aa,Delgado:2014vha,Delgado:2014kqa}.

   \item
     States localized in the brane, i.e.~third generation of squarks
     and third generation of slepton doublet, are naturally light as
     their tree-level masses are vanishing.  Their one-loop radiative
     masses from KK modes are finite~\cite{Delgado:1998qr} and can be
     interpreted as finite threshold effects after integrating out the
     heavy modes. Moreover left-handed and right-handed squarks do not
     mix much as their mixings are generated only at one loop as
     well. The values of the stop mixing $A_t$ and the one-loop masses
     of the fields localized in the brane are displayed in Fig.~\ref{massesrad}
     (their explicit expressions are given in Ref.~\cite{Delgado:1998qr}).
\begin{figure}[htb!]
\begin{center}
  \includegraphics[width=6.7cm]{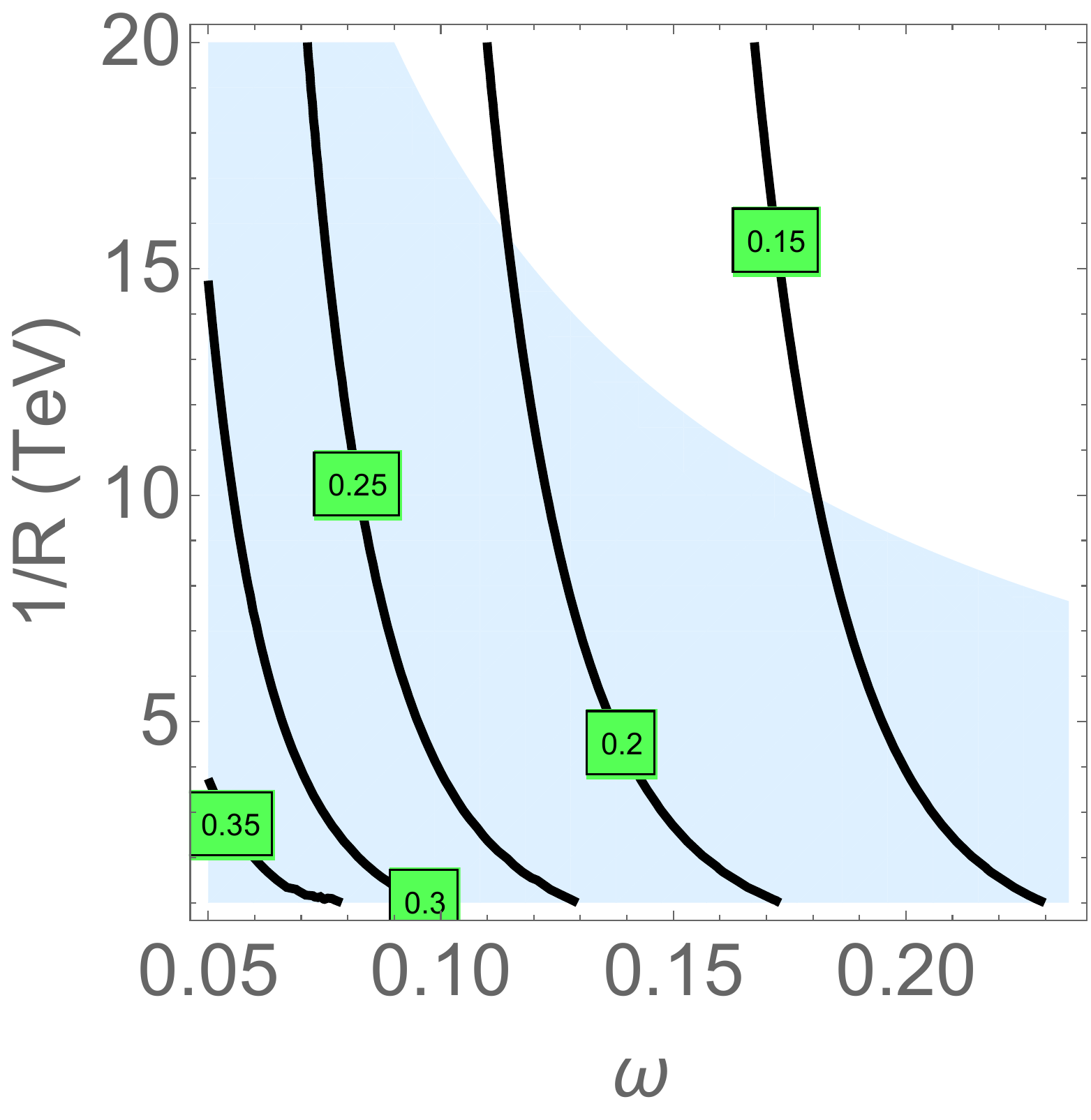} \hspace{0.7cm}
  \includegraphics[width=6.7cm]{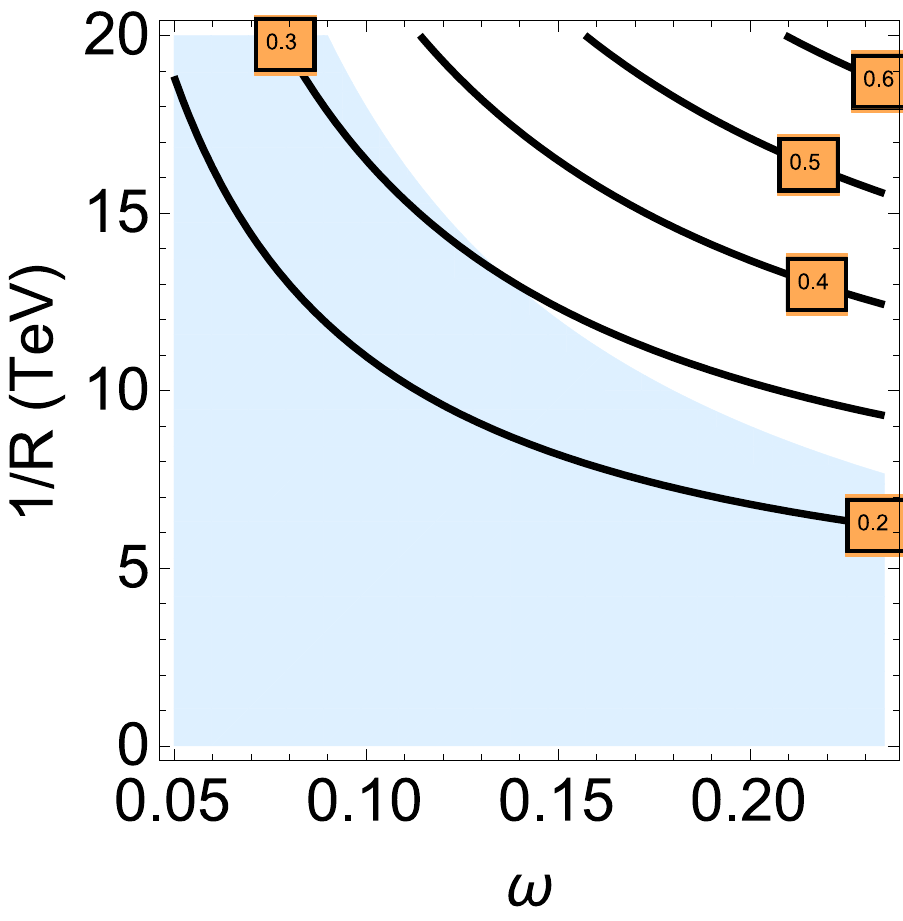} \\[2mm]
  \includegraphics[width=6.7cm]{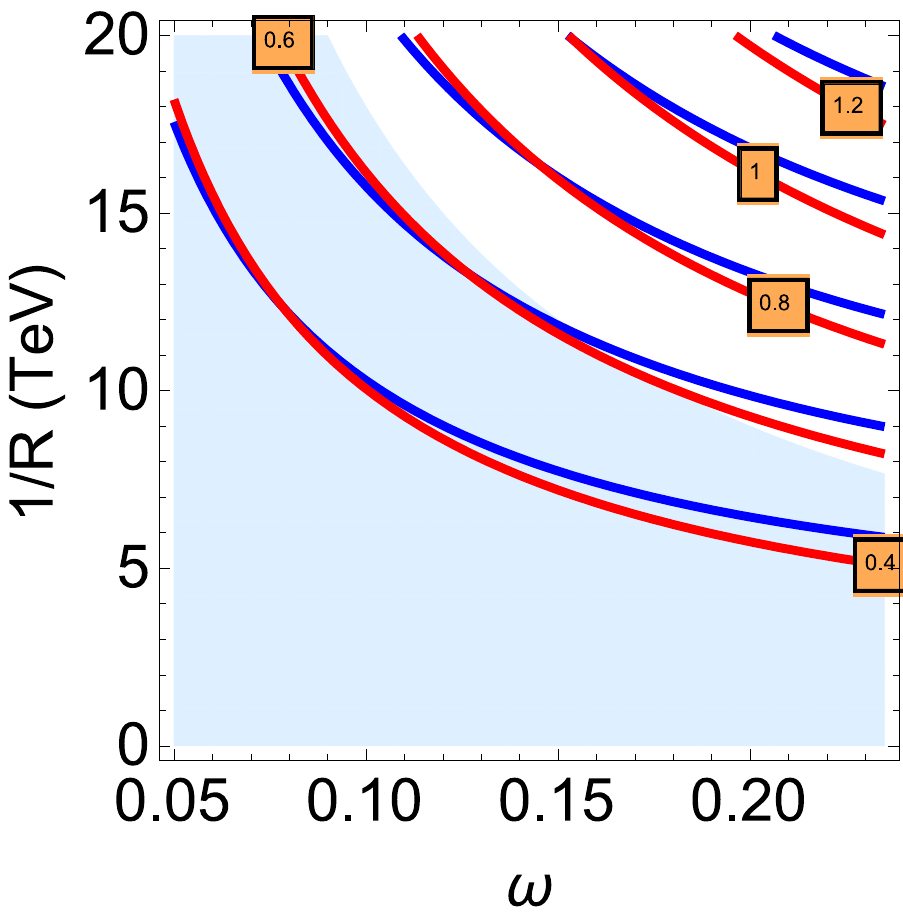} \hspace{0.7cm}
  \includegraphics[width=6.7cm]{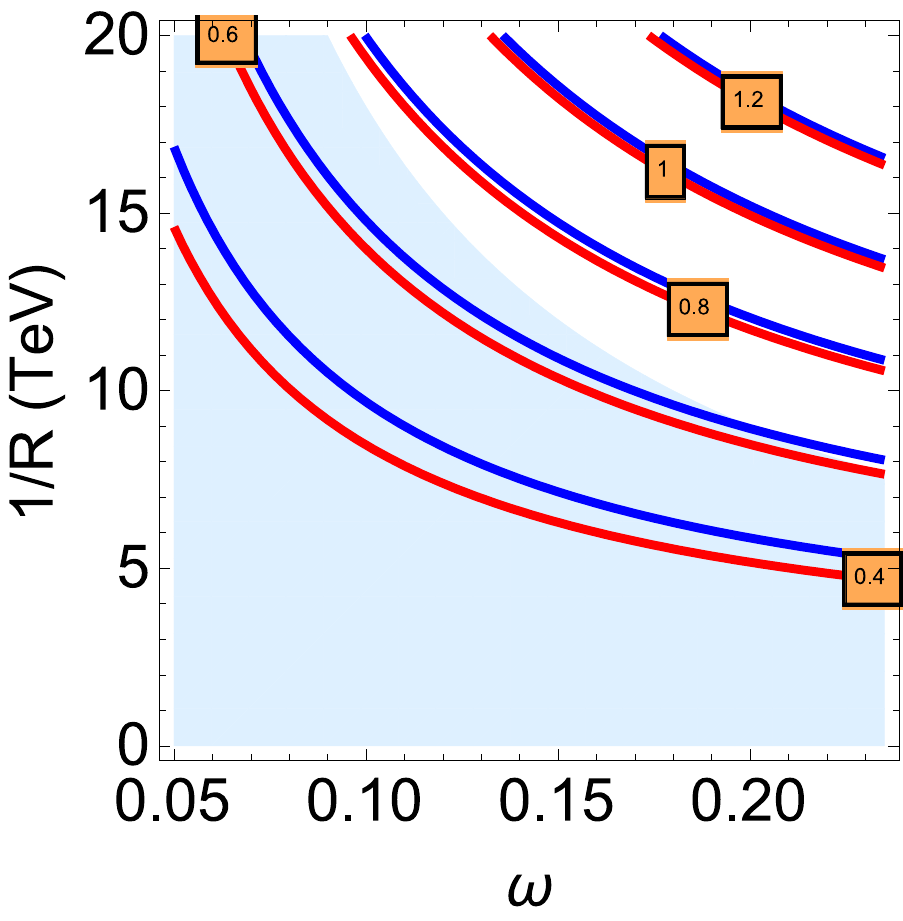}
  \caption{ \it Contour plots of the most relevant loop-induced
    parameters. In light blue the region with gluino mass
    $m_{\tilde{g}}<1.8$ TeV, in tension with LHC bounds (see
    Sec.~\ref{sec:low}).  Mass labels are in TeV units.  Upper left
    panel: Stop trilinear parameter normalized as $X_t=A_t/m_Q$. Upper
    right panel: Masses of the scalar left-handed tau $\tilde
    \tau_L$ and the scalar left-handed tau neutrino $\tilde \nu_{\tau}$. Lower left panel:
    Masses of the lightest states of the stop $\tilde t_1$ and sbottom
    $\tilde b_1$ (red and blue lines respectively). Lower right panel:
    Masses of the heaviest states of the stop $\tilde t_2$ and sbottom
    $\tilde b_2$ (red and blue lines respectively).}
\label{massesrad}
\end{center}
\end{figure}
\item
     The lightest $(n=0)$ modes of the fields in the bulk have tree-level
     masses that are zero, $\omega/R$, $2\omega/R$ or
     $1/R$. Those with vanishing masses correspond to SM-like fields. The
     new-physics spectrum thus exhibits very compressed sectors, with
     a large gap between new-physics bulk and brane states.  In this
     way the first and second generation sfermions, and the
     right-handed staus as well, are naturally much heavier than the
     stops, sbottoms, and left-handed staus and tau sneutrinos, in
     agreement with flavor constraints. The explicit values of the
     lightest new-physics modes are presented in
     Fig.~\ref{massestree}.
\begin{figure}[h!]
\begin{center}
 \includegraphics[width=7.cm]{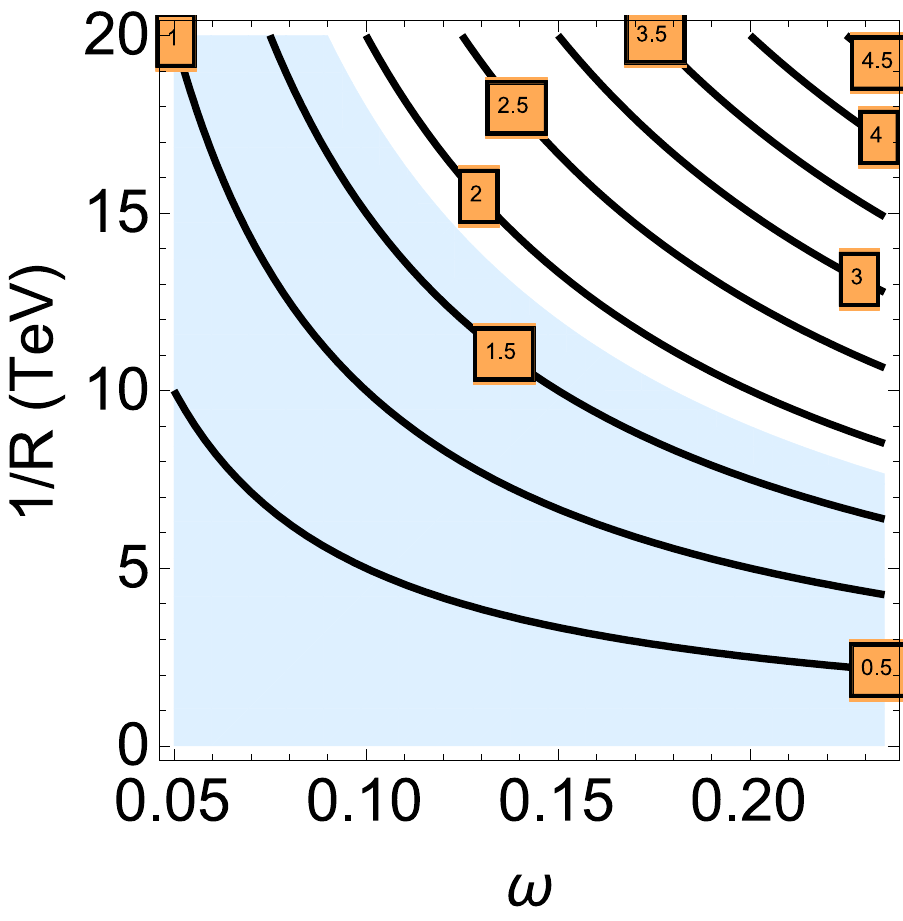}  \hspace{0.7cm}
  \includegraphics[width=7.cm]{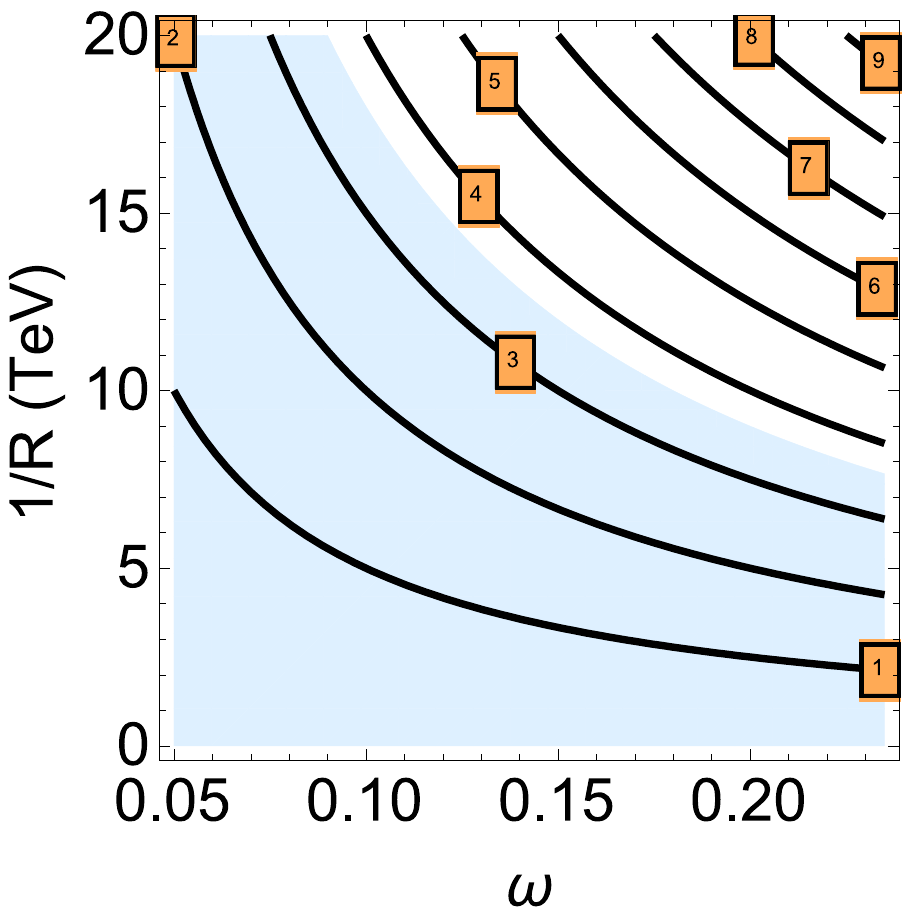}

  \caption{\it Contour plots of the tree-level masses of the bulk
    fields sensitive to the SS mechanism. Labels are in TeV units.
    Left panel: First and second generation sfermions, right-handed
    stau, gauginos and Higgsinos. Right panel: Charged and neutral
    heavy doublet Higgses. Blue areas are as in Fig.~\ref{massesrad}.}
\label{massestree}
\end{center}
\end{figure}

   \item
     The EWSB has to proceed by radiative corrections as discussed
     here below.

\end{itemize}
The minimal picture suffers then from two drawbacks:
 \begin{description}
 \item[i) EWSB:] In this theory the radiative corrections to the mass
   terms are
   known~\cite{Pomarol:1998sd,Antoniadis:1998sd,Delgado:1998qr}. They
   are finite and can be considered as threshold effects at the
   compactification scale $\mathcal O(1/R)$ at which all heavy bulk
   fields are integrated out. In particular, at one loop, there are
   gauge corrections to the squared mass of the SM Higgs ($m^2$) and
   the brane fields $\widetilde Q_{3L},\, \widetilde U_{3R},\,
   \widetilde D_{3R},\, \widetilde E_{3R}$,
   $(m_Q^2,m_U^2,m_D^2,m_E^2)$, which are positive, thus preventing
   EWSB. As stops are localized, and massless at tree level, they do
   not produce any {\it one-loop} correction to the Higgs mass
   proportional to $h_t^2$ which could trigger EWSB as in the 4D
   MSSM. Of course, when they are integrated out, they generate a
   (logarithmic) radiative correction depending on their own
   (one-loop) masses: a two-loop effect. In the $\overline{MS}$ scheme
   this correction is given by~\cite{Masina:2015ixa}
\begin{align}
\Delta m^2&=\frac{6h_t^2}{32\pi^2}\left[G(m_Q^2)+G(m_{U}^2)\right]+6h_t^2A_t^2\frac{G(m_Q^2)-G(m_U^2)}{m_Q^2-m_U^2}\, ,\nonumber\\
G(x)&\equiv x^2\left(\log\frac{x^2}{\mathcal Q^2}-1\right)\, .
\label{previouslog}
\end{align}
To leading order in $\alpha_3$ it turns out that  $m_Q(\omega)=m_U(\omega)$
and
Eq.~\eqref{previouslog} becomes
\be
\Delta m^2=-\frac{3h_t^2}{8\pi^2}m_Q^2(\omega)\left( \log\frac{\mathcal Q^2}{m_Q^2}+1  \right)\, .
\label{deltalog}
\ee
In Eq.~\eqref{deltalog} we can set the renormalization scale $\mathcal
Q$ at the scale where the boundary conditions are imposed, i.e.~where
the theory becomes 4D. In Ref.~\cite{Pomarol:1998sd} it was taken as
$\mathcal Q\simeq\omega/R$ whereas in Ref.~\cite{Dimopoulos:2014aua}
it was fixed as $\mathcal Q\simeq 1/(\pi R)$. In both cases the
two loop correction coming from $m_{Q}^2$, $m_{U}^2$, and $A_t^2$,
which are generated only at one loop, are too small to drive $m^2<0$.
On the other hand, the choice of $\mathcal Q$ in Eq.~(\ref{deltalog})
only concerns the scale dependence in the three loop
contribution~\footnote{Note that the EWSB could be determined by means
  of the RG-improved effective potential.  In that case the resulting
  EWSB condition is independent of $\mathcal Q$ at a given
  perturbative order~\cite{Masina:2015ixa}. The choice of $\mathcal Q$
  is then only aimed to minimize the corrections coming at the next
  perturbative order.}. Since we are not calculating all consistent
two loop effects (e.g.~we integrate out the heavy KK modes only at
one loop) our EWSB analysis should not rely on just the few two loop
pieces that are known, and which do not to change either
qualitatively or quantitatively the EWSB picture, as we have
checked. Thus, to be consistent, we consider the EWSB at one loop and
hereafter we will then ignore all two-loop EWSB contributions as the
one in Eq.~(\ref{deltalog}).

\item[ii) Higgs mass:] As both stop soft masses and trilinear stop mixing parameter
  are one-loop suppressed, their radiative correction to the Higgs
  quartic coupling is too small to reproduce the experimental value
  $m_h\simeq 125\,$GeV.  This problem was already recognized in the
  early papers~\cite{Pomarol:1998sd,Antoniadis:1998sd,Delgado:1998qr}
  and has been more recently
  revamped~\cite{Dimopoulos:2014aua,Garcia:2015sfa}.
 \end{description}

In Ref.~\cite{Dimopoulos:2014aua} problem {\bf (i)} was solved by
the introduction of higher dimensional operators, while problem {\bf (ii)} was
milder; as for the maximal SS breaking case $\omega=1/2$ the EOM lead
to $\tan\beta=\infty$, and are solved by introducing an extra $U(1)$
factor. In Ref.~\cite{Cohen:2015gaa} both problems were addressed by
adding a singlet plus a folded sector (i.e.~a copy of the matter
superfields) at the expense of bilinear mass term
parameters in the superpotential.  In the present paper we will see
that an extended Higgs sector can solve both problems without
violating the requirements {\it (i), (ii)} and $(iii)$ of
Sec.~\ref{sec:intro}. As an extra singlet is appropriate to add
tree-level corrections (an $F$-term contribution) to the Higgs mass in the case of $\tan\beta=1$,
we will first start considering the case of a localized singlet.
 
\section{5D MSSM plus a singlet}
\label{singlets}
Our setup in this section will be identical to that of Sec.~\ref{MSSM}  but with the addition of a singlet. For simplicity we will first consider the case of a singlet field $S$ localized on the $y=0$ brane.

\subsection{Embedding and 4D Lagrangian}
A singlet localized on the $y=0$ brane admits
a superpotential interaction with the (bulk) Higgs
multiplets that can be derived from the brane superpotential
\be W=\widehat\lambda \mathcal H_1\cdot \mathcal H_2S\delta(y)~,
\label{eq:superpotS}
\ee
where $\widehat\lambda$ is a 5D Yukawa coupling with mass dimension
equal to $-1$.  Specifically, only the even Higgs components couple to
the fields on the $y=0$ brane, so that the corresponding $N=1$ superfields $\mathcal H_1$ and $\mathcal
H_2$ are given by~\cite{Mirabelli:1997aj}
 \begin{align}
 \mathcal H_2&=(H_2,\psi_{2L},F_2-\partial_5 H_2^c)\nonumber~, \\
 \mathcal H_1&=(\widetilde H_1,\widetilde{\psi}_{1R},\widetilde F_1-\partial_5 \widetilde H_1^{c})~ ,
 \label{eq:H1H2}
 \end{align}
where, for a doublet $A$ with hypercharge 1/2, $\widetilde
A=-i\sigma_2 A^\star$ stands for a doublet with hypercharge $-1/2$.  In
particular, the fermionic components of $\mathcal H_{1,2}$ interact
with the singlet as a Dirac fermion $\Psi$ defined as
(cf.~Eq.~\eqref{eq:Hinos})
\be
\Psi=\sum_{n=0}^\infty\psi^{(n)}\equiv \sum_{n=0}^\infty\left( \begin{matrix} \psi_{2L}^{(n)}\\ \psi_{1R}^{(n)} \end{matrix}  \right)=
\widetilde H^{(0)}+\frac{1}{\sqrt{2}}\sum_{n=1}^\infty\left(
\widetilde H^{(n)} + \widetilde H^{(-n)} \right)
\equiv\frac{1}{\sqrt{2}}\widetilde H~.
\ee

In fact, from Eqs.~\eqref{eq:superpotS} and \eqref{eq:H1H2} one can determine
the 4D Lagrangian. After integrating out the auxiliary fields, its bosonic part reads
\begin{align}
 \mathcal L_{4}=&-\widehat\lambda S \Big\{\partial_5 H_1^{c\dagger}(0) H_2(0)+H_1^\dagger(0) \partial_5H_2^c(0)   +h.c. \Big\}\nonumber\\
 &-\widehat\lambda^2\Big\{ |H_1(0)^\dagger H_2(0)|^2 +|S|^2\left(|H_1(0)|^2+|H_2(0)|^2\right)\pi\delta(0)\Big\}\, ,%
  \label{4DD} 
\end{align}
with
\be
\delta(0)\equiv \frac{1}{\pi}\sum_{n=-\infty}^\infty 1\, .
\label{eq:delta}
\ee
Moreover, using the notation
\bea
h=\sum_{n=-\infty}^\infty h^{(n)}~,& \quad &\widehat h=\sum_{n=-\infty}^\infty (q_R-q_H+n)h^{(n)}~, \\
H=\sum_{n=-\infty}^\infty H^{(n)}~,& \quad &\widehat H=\sum_{n=-\infty}^\infty (q_R+q_H+n)H^{(n)}~ ,
\eea
then
\bea
\partial_5H_1^c(0)=\frac{-1}{\sqrt{2\pi}}(\widehat h+\widehat H),\qquad  H_1(0)= \frac{1}{\sqrt{2\pi}}(h+H)~,\nonumber\\
\partial_5H_2^c(0)=\frac{1}{\sqrt{2\pi}}(\widehat h-\widehat H),\qquad   H_2(0)= \frac{1}{\sqrt{2\pi}}(h-H)~, 
\label{termsdoublets}
\eea
and Eq.~(\ref{4DD}) reads
\begin{align}
\mathcal L_4=&-\frac{\lambda}{2} \Big\{S(h^\dagger+H^\dagger)(\widehat h-\widehat H)-S^\dagger(h^\dagger-H^\dagger)(\widehat h+\widehat H)  +h.c.\Big\}\nonumber\\
&-\lambda^2 \Big\{ \frac{1}{4}\left| |h|^2-|H|^2-h^\dagger H+H^\dagger h   \right|^2 
+|S|^2(|h|^2+|H|^2)\pi\delta(0) \Big\}\, ,
\label{4Dsinglete}
\end{align}
where $\lambda\equiv\widehat\lambda/\pi$ is the (dimensionless)
4D Yukawa coupling.

\subsection{Quartic and quadratic terms of the lightest Higgs}
\label{sec:quarticSing}
As we can see from Eq.~(\ref{4Dsinglete}), the coupling $\lambda$ is
the tree level source of the $h^{(0)}$ quartic coupling.  The $h^{(0)}$
potential is then given by
\be
V_{SM}=(m^2+\Delta_h m^2) |h^{(0)}|^2+\left(\frac{\lambda^2}{4}+\Delta\lambda\right)|h^{(0)}|^4+\dots~.
\label{eq:SMpot}
\ee
where $m^2$ is the tree-level Higgs squared mass term while $\Delta_h
m^2$ and $\Delta \lambda$ are the radiative contributions to the Higgs
mass and quartic coupling, respectively.

\begin{figure}[h!]
\begin{center}
 \includegraphics[width=7.5cm]{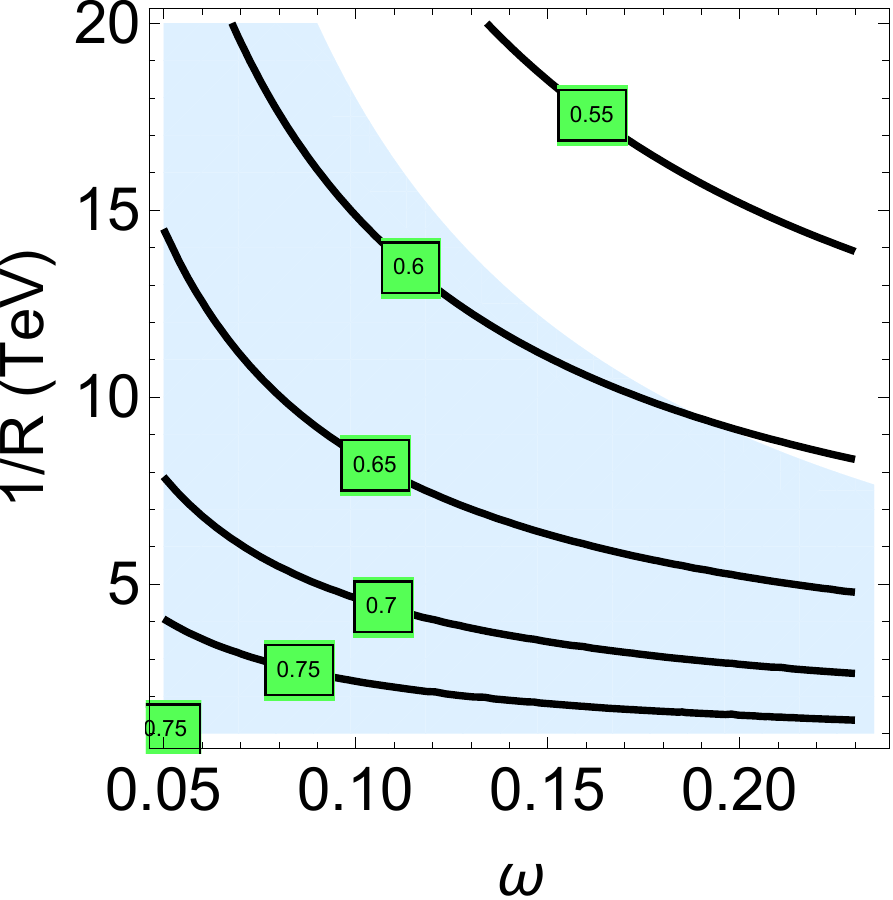}  \hspace{0.7cm}
  \includegraphics[width=7.5cm]{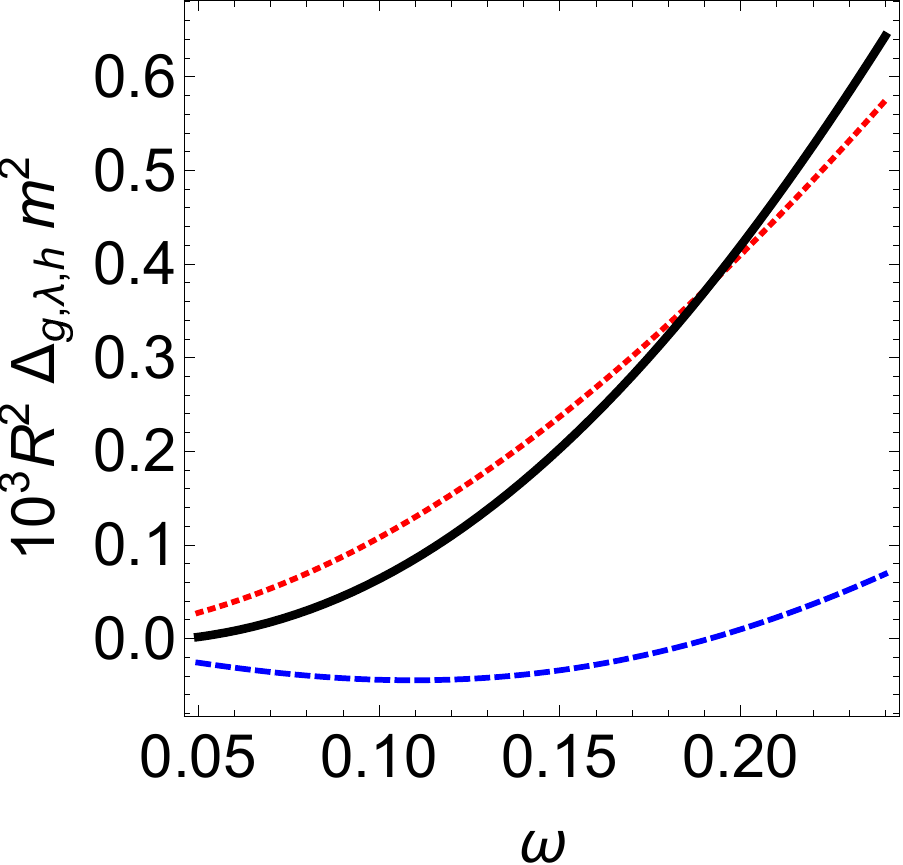}
  \caption{\it Left panel: Contour plot of values of $\lambda$ leading
    to the experimental value $m_h=125$\,GeV if the observed EWSB is
    achieved. Blue area is as in Fig.~\ref{massesrad}. Right panel: Plot
    of $10^3R^2\Delta_g m^2$ (red, dotted line),
    $10^3R^2\Delta_\lambda m^2$ (blue, dashed line) and its sum
    $10^3R^2\Delta_h m^2$ (black, solid line), for $1/R=2$ TeV and
    $\lambda$ fixed from the plot on the left panel, as a function of
    $\omega$.}
\label{plot:singlete}
\end{center}
\end{figure}

We determine the total $h^{(0)}$ quartic coupling at
leading order in $\lambda$ and $h_t$. Since the $\lambda$ dependence
already appears at tree level, $\Delta \lambda$ is the usual MSSM
radiative correction~\cite{Carena:1995bx}
\be
\Delta\lambda=\frac{3 m_t^4}{8\pi^2 v^4}\left[\log\frac{m^2_{\tilde t}}{m_t^2}+\frac{A_t^2}{m_{\tilde t}^2}\left(1-\frac{A_t^2}{12 m_{\tilde t}^2}  \right)        \right]\, ,
\label{eq:lamMSSM}
\ee
in which $v=174\,$GeV (i.e.~where the observed EWSB is assumed).  Notice
that as both $m_{\tilde t}^2\simeq m_U^2 \simeq m_Q^2$ and $A_t$ are
generated at one loop by exchange of KK modes (see
e.g.~Ref.~\cite{Pomarol:1998sd}), $\Delta \lambda$ is a two loop
effect.
Moreover, if we assume $m^2+\Delta_h m^2\simeq -(88\, \rm GeV)^2$, in
agreement with the EWSB observations, Eqs.~\eqref{eq:SMpot} and
\eqref{eq:lamMSSM} can be used to translate the constraint on the
$h^{(0)}$ scalar mass, $m_h$, into $\lambda$. This is quantified in
Fig.~\ref{plot:singlete} (left panel) where the explicit values of
$\lambda$ providing $m_h=125$\,GeV are displayed as a function of
$1/R$ and $\omega$, with the correction $\Delta \lambda$ being
included.

\begin{figure}[htb]
\begin{center}
\vspace{-.5cm}
\includegraphics[width=11cm,angle=270]{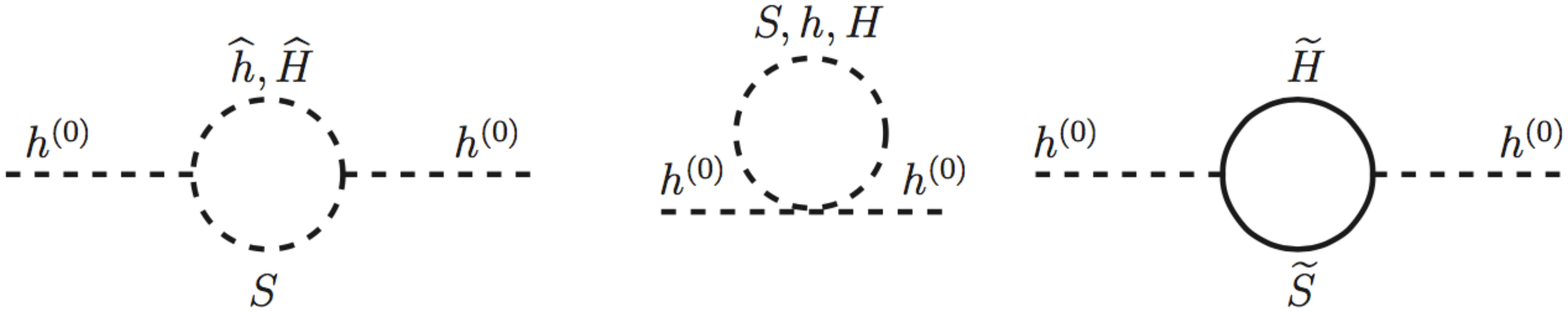}
\vspace{-4cm}
\caption{\it Diagrams contributing to the 
  correction $\Delta_\lambda m^2 $ to the squared-mass term of $h^{(0)}$.}
\label{singletdiagrams}
\end{center}
\end{figure}

On the other hand, it is not obvious that the EWSB condition
$m^2+\Delta_h m^2\simeq -(88\, \rm GeV)^2$ can be fulfilled.  As
discussed in Sec.~\ref{MSSM}, $m^2$ is vanishing. The EWSB then
relies only on the loop-induced quantity $\Delta_h m^2$. This can be
split as $\Delta_h m^2=\Delta_g m^2+\Delta_\lambda m^2$, where $\Delta_g
m^2$ and $\Delta_\lambda m^2$ are the contributions depending,
respectively, on the $SU(2)_L$ gauge coupling $g$ and on the
superpotential parameter $\lambda$~\footnote{For simplicity we are neglecting here the subleading contribution corresponding to the $U(1)_Y$ gauge interactions.}. The quantity $\Delta_g m^2$
amounts to~\cite{Pomarol:1998sd}
\begin{align}
  \Delta_gm^2&=\frac{g^2}{64\pi^4}\left[9\,\Omega (0)+3\,\Omega(2\omega)-12\,\Omega(\omega)   \right]\,
  \label{eq:Deltag}
\end{align}
with 
\be
\Omega(\omega)=\frac{1}{2}\big[\Li_3(e^{2i\pi \omega})+\Li_3(e^{-2i\pi \omega})\big](1/R)^2~.
\ee
The contribution $\Delta_\lambda m^2$ is generated by the diagrams in
Fig.~\ref{singletdiagrams} and turns out to be
\begin{equation}
\Delta_\lambda m^2(\omega)=\frac{\lambda^2}{32\pi^4}\left[ \Omega(0)+\Omega(2\omega)-2\,\Omega(\omega)\right]~.
\end{equation}

Plots of $\Delta_gm^2$, $\Delta_\lambda m^2$ and $\Delta_h m^2$ as a
function of $\omega$ are shown in the right panel of
Fig.~\ref{plot:singlete}. In the plots the value of $\lambda$ is adjusted
to reproduce the Higgs mass constraint (cf.~left panel of
Fig.~\ref{plot:singlete}) assuming that EWSB occurs. A representative value of $R$ is assumed,
namely $1/R=2\,$TeV. For this illustrative case it results that
$\Delta_\lambda m^2$, although negative for $0\lesssim\omega\lesssim
0.2$, is insufficient to overcome the positive contribution $\Delta_g
m^2$ and drive $\Delta_h m^2$ to negative values. We check that this
negative result is generic.

We conclude that in the 5D MSSM plus a localized singlet, the extra
field content helps in reproducing the experimental value of the Higgs
mass but does not seem to improve the scenario from the EWSB problem.
A possible solution is to introduce higher-dimensional operators as in
Ref.~\cite{Dimopoulos:2014aua}. However, a subtlety in the analysis
might be exploited to circumvent the problem: if there is sizable
mixing between the singlet and Higgs scalars, $\Delta_h m^2$ is not the
unique quantity playing a role in the EWSB. We sketch the features of
this possibility in the following section.

\subsection{Tadpole and VEV of the singlet}

\begin{figure}[h!]
\vspace{-1.5cm}
\begin{center}
\includegraphics[width=7.cm]{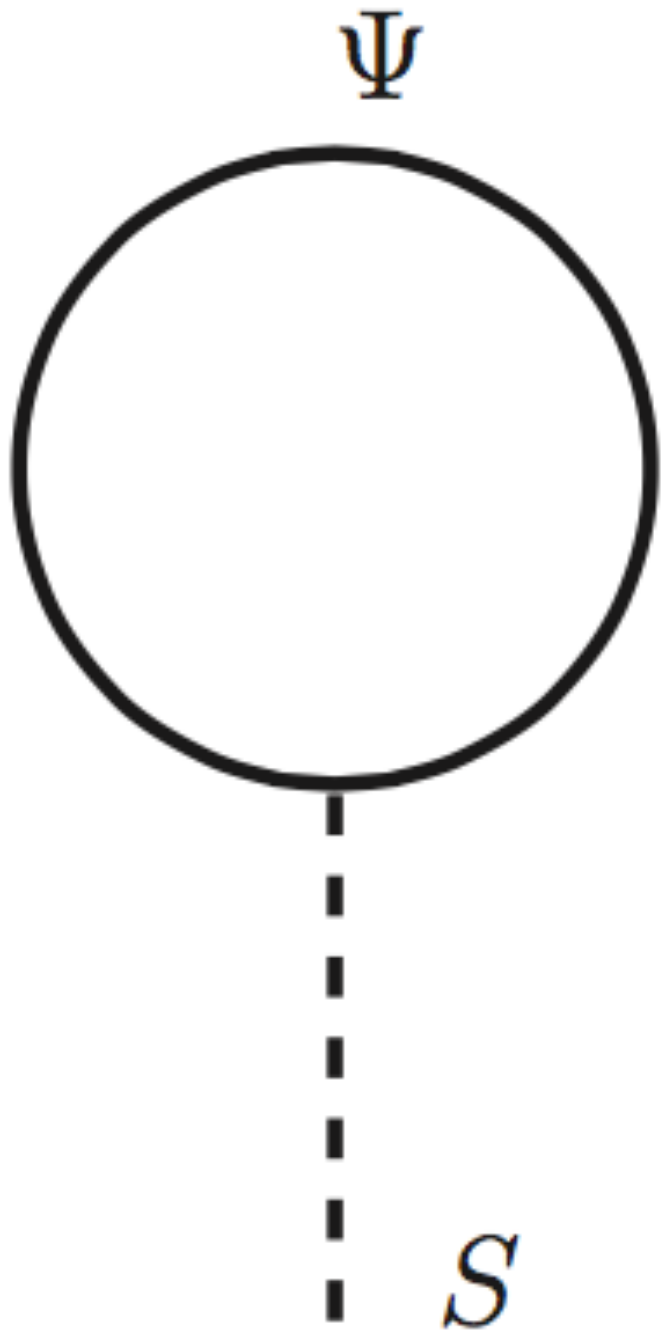}\hspace{0.5cm}
 \includegraphics[width=7.5cm]{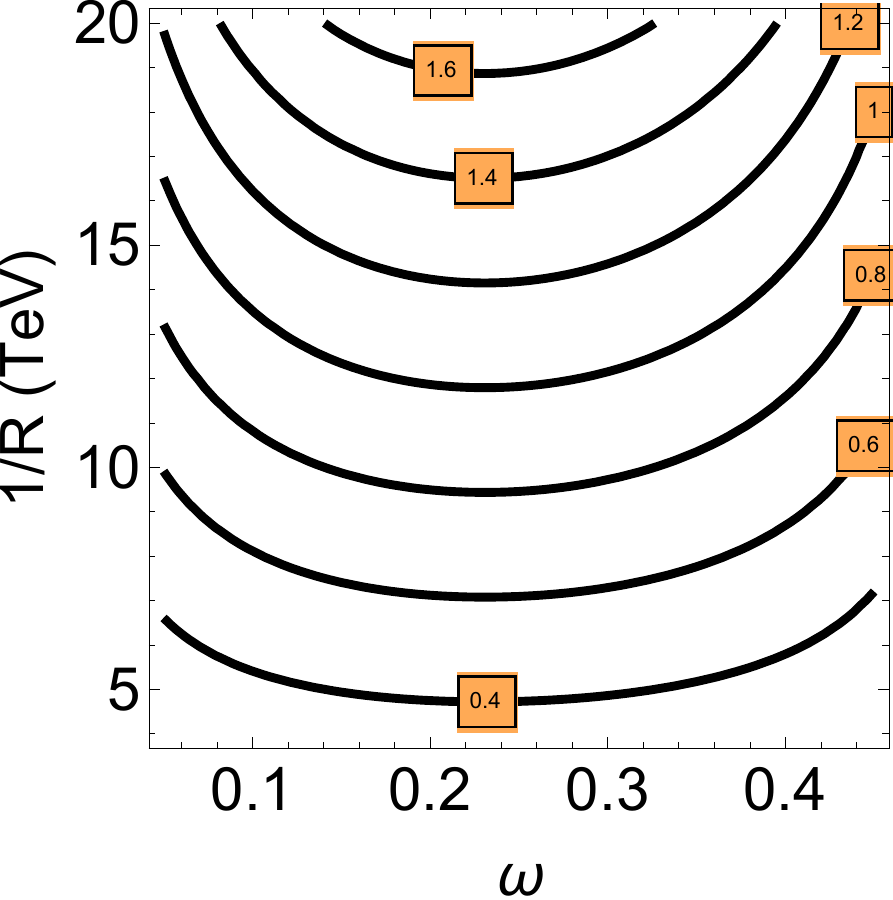} 
\vspace{-0.2cm}
\caption{\it Left panel: Diagrams contributing to the tadpole of
  $S$. Right panel: Contour plot of the triplet trilinear parameter
  $\xi^{1/3}(\omega,1/R)$ with $\lambda$ adjusted to reproduce the
  observed Higgs mass. Labels are in TeV units.}
\label{diagram:tadpole}
\end{center}
\end{figure}

The interaction between the singlet and the fermions of $\mathcal
H_{1,2}$ allows the Feynman diagram of Fig.~\ref{diagram:tadpole}
(left panel) to generate the linear term in the Lagrangian
proportional to $S$, as it is not protected by any symmetry of the
theory. Then there exists a tadpole term in the localized Lagrangian
as
\be
\mathcal L_{4}=-\xi(\omega,1/R) (S+S^\dagger)+\dots\,~ .
\ee
The coefficient of this interaction is expected to be
sizable~\footnote{This contribution has not been noticed in the
  previous literature. The recent
  proposals~\cite{Garcia:2015sfa,Cohen:2015gaa} might be affected by
  this tadpole term.}. Indeed, the sum of the contributions of each KK
mode $\psi^{(n)}$ yields
\be
\xi(\omega,1/R)=\frac{3i}{32\pi^5}\left[\Li_4\left(e^{-2i\pi\omega}\right)-\Li_4\left(e^{2i\pi\omega}\right)\right]\left(1/R\right)^3\, ,
\ee
and its numerical value can be deduced from the right panel of
Fig.~\ref{diagram:tadpole}. The size $\mathcal O(0.1/R)$ is then
expected to be the natural scale of the dimensionful parameters
involved in the singlet potential, so that the VEV that is eventually
acquired by the singlet should be parametrically $\mathcal O(1$\,TeV)
for $1/R\sim 10\,$TeV. The fields $h^{(0)}$ and $S$ can thus have a
non-negligible mixing. In principle the mixing could help the
implementation of the EWSB at the expense of some tension with the
Higgs signal strengths measured at the
LHC~\cite{Khachatryan:2016vau}~\footnote{Although $h^{(0)}$ and $S$
  have positive (one-loop) squared-mass terms, there may be a linear
  combination developing a negative quadratic term thanks to the
  mixing.}.  Determining whether these possibilities are not ruled out
by the present LHC data would need a dedicated analysis that goes
beyond the scope of the present paper, and in any case we do not
expect the surviving parameter region to be really promising
concerning naturalness. On the other hand, the situation would not
radically change by considering singlets in the bulk, as bulk singlets
still acquire a large VEV. We thus focus the rest of our analysis on
the 5D MSSM extended by hyperchargeless $SU(2)_L$ triplets for which
tadpoles prior to the EWSB are forbidden by the gauge symmetry.

\section{5D MSSM plus bulk triplets}
\label{sec:triplets}

We consider the scenario where the Higgs sector is extended by
hyperchargeless $SU(2)_L$ triplets. In the context of 4D supersymmetry
the model is somewhat well known (see
e.g.~Refs.~\cite{Espinosa:1992hp, Espinosa:1991wt,
  DiChiara:2008rg,Delgado:2012sm,Delgado:2013zfa,Bandyopadhyay:2013lca,
  Arina:2014xya, Bandyopadhyay:2014vma}) but its implementation in a
5D framework has not been attempted yet. In this section we implement
it in a SS scenario. As we refrain from introducing any dimensionful
parameter in the 4D superpotential, we do not consider the option of
triplets localized on the brane (in which case the fermionic triplet components
would be too light to overcome the chargino mass bound $m_{\widetilde
  \chi^\pm}\gtrsim 104\,$ GeV~\cite{Heister:2002mn}).
We thus consider the 5D MSSM extensions with triplets propagating in
the bulk.

\subsection{Embedding and 4D Lagrangian}

Similar to the case of bulk doublets (see Sec.~\ref{MSSM}), the
bulk triplets can be arranged in $Y=0$ $SU(2)_L$-triplet
hypermultiplets $\mathbb
T_b=(\Sigma_b,\Sigma_b^c,\Psi_{\Sigma_b},F_{\Sigma_b},F_{\Sigma_b})$
with $b=1,2$, which transform as a doublet under the global bulk group $SU(2)_\Sigma$ acting on the index $b$. The fermionic component
$\Psi_{\Sigma_b}=(\psi_{\Sigma_b}, \bar\psi_{\Sigma_b}^c)^T$ is a
Dirac spinor while $\psi_{\Sigma_b}$ ($\bar\psi_{\Sigma_b}$) and
$\psi^c_{\Sigma_b}$ ($\bar\psi^c_{\Sigma_b}$) are undotted (dotted)
Weyl spinors.  Concerning the $\mathbb Z_2$ symmetry, we assume the
multiplets $(\Sigma_2,\psi_{\Sigma_2},F_{\Sigma_2})$ and
$(\Sigma_1,\bar\psi_{\Sigma_1}^c,F_{\Sigma_1)}$ to be even and
$(\Sigma_1^c,\psi_{\Sigma_1},F_{\Sigma_1}^c)$ and
$(\Sigma_2^c,\bar\psi_{\Sigma_2}^c,F_2^c)$ to be odd, according to the orbifold action
\be
\mathbb Z_2=\left.\sigma_3\right|_{SU(2)_\Sigma}\otimes\gamma_5
\ee
where $\sigma_3$ is acting over $SU(2)_\Sigma$ indices and $\gamma_5$ over Dirac indices.
We denote their scalar KK modes as $\Sigma_{1,2}^{(n)}$ ($n\geq 0$) and
$\Sigma_{1,2}^{c(n)}$ ($n\geq 1$).

This allows one to
introduce the SS twists $(q_R,q_\Sigma)$ that establish the transformation
\be
 \left[\begin{array}{cc} \Sigma_1(x,y)& \Sigma_1^c(x,y)\\ \Sigma_2^c(x,y) & \Sigma_2(x,y) \end{array}\right]=e^{i q_\Sigma\sigma_2 y}
 \sum_{n=0}^\infty  \sqrt{\frac{2}{\pi}} \left[\begin{array}{cc} \cos ny\, \Sigma_1^{(n)}(x)& \sin ny\, \Sigma_1^{c(n)}(x)\\ \sin ny\, \Sigma_2^{c(n)}(x) & \cos ny\, \Sigma_2^{(n)}(x) \end{array}\right]
 e^{-i q_R\sigma_2 y}~,
 \label{triplets}
 \ee
whose mode normalization is in analogy with Eq.~\eqref{doublets}.  The
pattern of the mass eigenvalues and the spectrum of the triplet is also
similar to the ones in Eqs.~\eqref{nonceroH} and
\eqref{ceroH}. Indeed, applying the same normalization conventions, it
turns out that the mass eigenstates $\sigma^{(n)}$ and $\Sigma^{(n)}$,
with mass $q_R-q_\Sigma+n$ and $q_R+q_\Sigma+n$ respectively, are
given by
\begin{align}
\Sigma_1^{(n)}=&\left(\sigma^{(n)}+\sigma^{(-n)}+\Sigma^{(n)}+\Sigma^{(-n)}   \right)/2\, ,\nonumber\\
\Sigma_2^{(n)}=&\left(\sigma^{(n)}+\sigma^{(-n)}-\Sigma^{(n)}-\Sigma^{(-n)}   \right)/2\, ,\nonumber\\
\Sigma_1^{c(n)}=&\left(\sigma^{(-n)}-\sigma^{(n)}+\Sigma^{(-n)}-\Sigma^{(n)}   \right)/2\, ,\nonumber\\
\Sigma_2^{c(n)}=&\left(\sigma^{(n)}-\sigma^{(-n)}+\Sigma^{(-n)}-\Sigma^{(n)}   \right)/2\, ,
\label{nonceroS}
\end{align}
for $n\ge1$, and by 
\begin{align}
\Sigma_1^{(0)}=&\left( \sigma^{(0)}+\Sigma^{(0)}  \right)/2\, ,\nonumber\\
\Sigma_2^{(0)}=&\left( \sigma^{(0)}-\Sigma^{(0)}  \right)/2\, ,
\label{ceroS}
\end{align}
for $n=0$. The analogy also applies to the fermionic components of the
triplet. Their tree-level mass spectrum is then similar to the one of the
Higgsinos.

Only the even multiplets can have interactions on the $y=0$ brane, and
the $N=1$ triplet supermultiplets that have such interactions are
 \begin{align}
 \mathcal T_2&=(\Sigma_2,\psi_{\Sigma_2},F_{\Sigma_2}-\partial_5 \Sigma_2^c)\, ,\nonumber\\
 \mathcal T_1&=(\Sigma_1^\dagger,\psi_{\Sigma_1}^c,F_{\Sigma_1}^\dagger-\partial_5 \Sigma_1^{c\dagger})\, .
 \end{align}
The generic brane superpotential involving these fields is
 \be
 W=\left(\widehat\lambda_1 \mathcal H_1\cdot\mathcal T_1\mathcal H_2+\widehat\lambda_2 \mathcal H_1\cdot\mathcal T_2\mathcal H_2\,\right)\delta(y)\, ,
 \ee
where $\widehat\lambda_b$ are 5D Yukawa couplings with mass dimension
equal to $-3/2$. In particular, in the superpotential no triplet
tadpole or cubic terms are allowed by the gauge symmetry.

After integrating out the auxiliary fields we obtain the bosonic 4D
 Lagrangian
 \begin{align}
   \mathcal L_4=&-\Big\{H_1^\dagger(0)\left(\widehat\lambda_1\partial_5 \Sigma_1^{c\dagger}(0)+\widehat\lambda_2\partial_5 \Sigma_2^c(0)\right)H_2(0)
   +\partial_5 H_1^{c\dagger}(0) \left(\widehat\lambda_1 \Sigma_1^\dagger(0)+\widehat\lambda_2 \Sigma_2(0)\right)H_2(0)
   \nonumber \\
   &~~~~~+H_1^\dagger(0) \left(\widehat\lambda_1 \Sigma_1^\dagger(0)+\widehat\lambda_2 \Sigma_2(0)\right)\partial_5H_2^c(0)+h.c.\Big\}\nonumber\\
   &-\Big\{\frac{1}{2}(\widehat\lambda_1^2+\widehat\lambda_2^2)\sum_A\left|H_1^\dagger(0)\tau_A H_2(0)\right|^2\nonumber\\
 &~~~~~+\left| \left(\widehat\lambda_1 \Sigma_1^\dagger(0)+\widehat\lambda_2 \Sigma_2(0)\right)H_2 (0)\right|^2+\left| H_1^\dagger(0)\left(\widehat\lambda_1 \Sigma_1^\dagger(0)+\widehat\lambda_2 \Sigma_2(0)\right)\right|^2
 \Big\}\,\pi\delta(0)\, ,
  \label{4D}
 \end{align}
where $\tau_A$ are the Pauli matrices used in the definition
$\Sigma_b\equiv\frac{1}{\sqrt{2}}T_A^b\tau_A$. By means of the
notation
\bea
\sigma=\sum_{n=-\infty}^\infty \sigma^{(n)}~,& \quad &\widehat \sigma=\sum_{n=-\infty}^\infty (q_R-q_\Sigma+n)\sigma^{(n)}~, \\
\Sigma=\sum_{n=-\infty}^\infty \Sigma^{(n)}~,& \quad &\widehat \Sigma=\sum_{n=-\infty}^\infty (q_R+q_\Sigma+n)\Sigma^{(n)}~ ,
\eea
and the identities of Eq.~\eqref{termsdoublets} (with obvious replacements
$h\to \sigma$ and $H \to \Sigma$), Eq.~(\ref{4D}) reads
\begin{align}
\mathcal L_{4}=&\Big\{\frac{1}{2\sqrt{2}}(\lambda_1-\lambda_2)\left[h^\dagger\widehat\sigma h-H^\dagger\widehat\sigma H+h^\dagger\widehat\Sigma H-H^\dagger \widehat\Sigma h\right.\nonumber\\
-&\left. \widehat H^\dagger(\sigma-\sigma^\dagger)h+\widehat h^\dagger(\Sigma+\Sigma^\dagger)h+\widehat h^\dagger(\sigma-\sigma^\dagger)H-\widehat H^\dagger(\Sigma+\Sigma^\dagger)H \right]\nonumber\\
+&\frac{1}{2\sqrt{2}}(\lambda_1+\lambda_2)\left[h^\dagger\widehat\Sigma h-H^\dagger\widehat\Sigma H+h^\dagger\widehat\sigma H-H^\dagger \widehat\sigma h\right.\nonumber\\
+&\left. \widehat H^\dagger(\sigma+\sigma^\dagger)H-\widehat h^\dagger(\Sigma-\Sigma^\dagger)H+\widehat h^\dagger(\sigma+\sigma^\dagger)h-\widehat H^\dagger(\Sigma-\Sigma^\dagger)h\right]+h.c.\Big\}\nonumber\\
-&\frac{1}{4}\left[ h^\dagger F_+ h+H^\dagger F_+ H+h^\dagger F_- H+H^\dagger F_- h \right] \pi \delta(0)\nonumber\\
-&\frac{\lambda_1^2+\lambda_2^2}{8}\left(|h|^4+|H|^4+6|h|^2|H|^2-6|h^\dagger H|^2
-(h^\dagger H)^2-(H^\dagger h)^2\right)\pi\delta(0)\, ,
\label{eq:L4bis}
\end{align}
where $\lambda_b\equiv\widehat\lambda_b/\sqrt{\pi^3}$ are the
dimensionless 4D Yukawa coupling  and $F_\pm$ is given by
\be
F_{\pm}=\lambda_1^2[\sigma^\dagger+\Sigma^\dagger,\sigma+\Sigma]_\pm +\lambda_2^2[\sigma-\Sigma,\sigma^\dagger-\Sigma^\dagger]_\pm+\lambda_1\lambda_2
\big\{ [\sigma-\Sigma,\sigma+\Sigma]_\pm+h.c.\big\}\ ,
\ee
with $[x,y]_-$ and $[x,y]_+$ standing for the commutator and
anticommutator operator, respectively. The decomposition
\be
\sigma\equiv\sum_A t_A \tau_A/\sqrt{2}
\label{eq:tA}
\ee
and the identity
$\tau^A_{ij}\tau^A_{k\ell}=2\delta_{i\ell}\delta_{jk}-\delta_{ij}\delta_{k\ell}$
have also been used.

\subsection{Quartic and quadratic terms of the lightest Higgs}
\label{sec:quartic}

\begin{figure}[htb]
\begin{center}
\vspace{-4cm}
\includegraphics[width=12cm,angle=270]{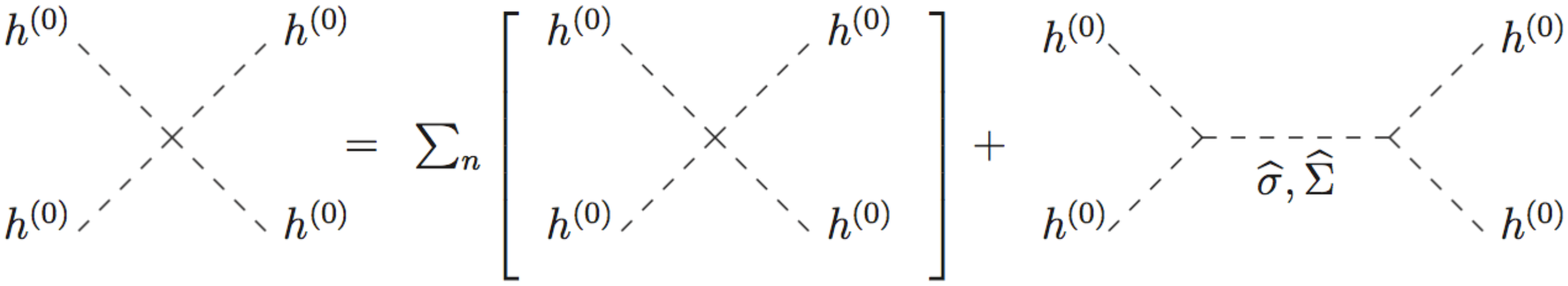}
 \vspace{-4cm}
 \caption{\label{fig:matchLamb} \it Matching of the low-energy and
   high-energy $h^{(0)}$ four-point diagrams.}
   
\end{center}
\end{figure}

From Eq.~\eqref{eq:L4bis} we can determine the potential of $h^{(0)}$
at low energy. On top of the contribution
$\pi\delta(0)(\lambda_1^2+\lambda_2^2)/8$, the low-energy quadratic
coupling includes the threshold correction due to the heavy modes that
are integrated out~(for a didactic calculation of threshold effects
see e.g.~\cite{Carena:2008rt,Draper:2016pys}). This relation is
provided by the tree-level matching condition depicted in
Fig.~\ref{fig:matchLamb}
where the
identity \eqref{eq:delta} has been used pictorially. For $q_R=q_H$ the
relation amounts to
\bea
  &&
  \lim_{p\to 0}\sum_{n=-\infty}^{+\infty}\frac{(\lambda_1+
  \lambda_2)^2}{16}\left[1+\frac{(q_R+ q_\Sigma+n)^2}{p^2-(q_R+
      q_\Sigma+n)^2}\right]|h^{(0)}|^4\nonumber\\
  &&+
   \lim_{p\to 0}\sum_{n=-\infty}^{+\infty}\frac{(\lambda_1-
  \lambda_2)^2}{16}\left[1+\frac{(q_R- q_\Sigma+n)^2}{p^2-(q_R-
      q_\Sigma+n)^2}\right]|h^{(0)}|^4\nonumber\\
  &&= \left[\frac{(\lambda_1+\lambda_2)^2}{16}\delta_{q_R+ q_\Sigma,0} +
   \frac{(\lambda_1-\lambda_2)^2}{16}\delta_{q_R- q_\Sigma,0}\right]|h^{(0)}|^4~.
  \label{lambda} 
\eea
For nonmaximal (and positive) twists the result is not vanishing only
if $q_R=q_\Sigma$ and the whole contribution is due to the $n=0$
mode. Therefore, in order to achieve a sizable boost to the tree-level
Higgs mass, we focus on the case $q_H=q_R=q_\Sigma\equiv\omega$ hereafter. The
low-energy potential of $h^{(0)}$ is then given by
\be
V_{SM}= (m^2+\Delta_h m^2) |h^{(0)}|^2+\left(\frac{(\lambda_1-\lambda_2)^2}{16}+\Delta\lambda\right)|h^{(0)}|^4+\dots ~.
\label{eq:Vtrip}
\ee

\begin{figure}[htb]
\begin{center}
 \includegraphics[width=7.5cm]{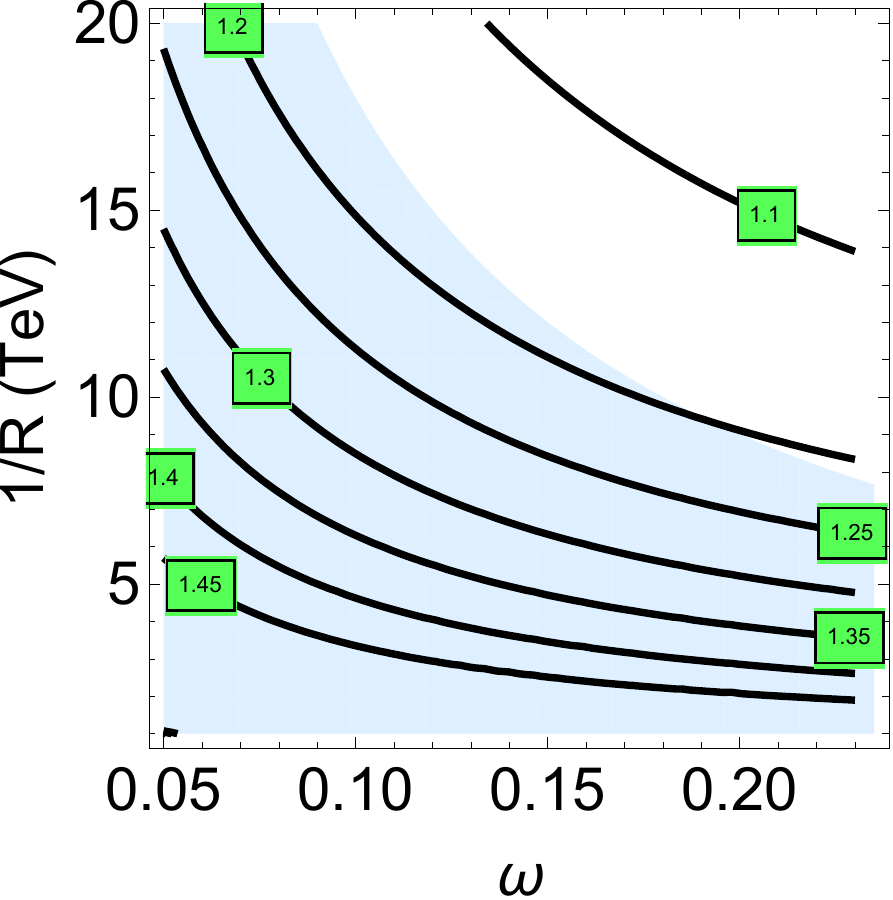} 
  \caption{\it Contour plot of values of $|\lambda_1-\lambda_2|$ fixing the Higgs mass to its experimental value $m_h=125$ GeV. Blue area is as in Fig.~\ref{massesrad}.
}
\label{lambdaminus}
\end{center}
\end{figure}

We then determine the $h^{(0)}$ quartic coupling at leading order in
$\lambda_{1,2}$ and $h_t$, as in previous sections. The contribution depending on
$\lambda_{1,2}$ appears at tree level while the latter appears at two
loop and is given by Eq.~\eqref{eq:lamMSSM} (see comments in
Sec.~\ref{sec:quarticSing}). Once the observed EWSB is assumed,
which in practice is equivalent to impose $m^2+\Delta_h m^2\simeq -(88\,
\rm GeV)^2$, the experimental measurement of the Higgs mass constrains
$|\lambda_1-\lambda_2|$, $R$ and $\omega$ as shown in
Fig.~\ref{lambdaminus}~\footnote{As we will discuss in
  Section~\ref{sec:TripVEV}, $h^{(0)}$ mixes very mildly with the
  scalar triplet. The field $h^{(0)}$ is then the mass eigenstate that
  plays the role of the SM-like Higgs. In addition, because of the
  small mixing, only the squared-mass term of $h^{(0)}$ is relevant in
  the EWSB.}.

\begin{figure}[htb]
\begin{center}
\vspace{-4cm}
\includegraphics[width=12cm,angle=270]{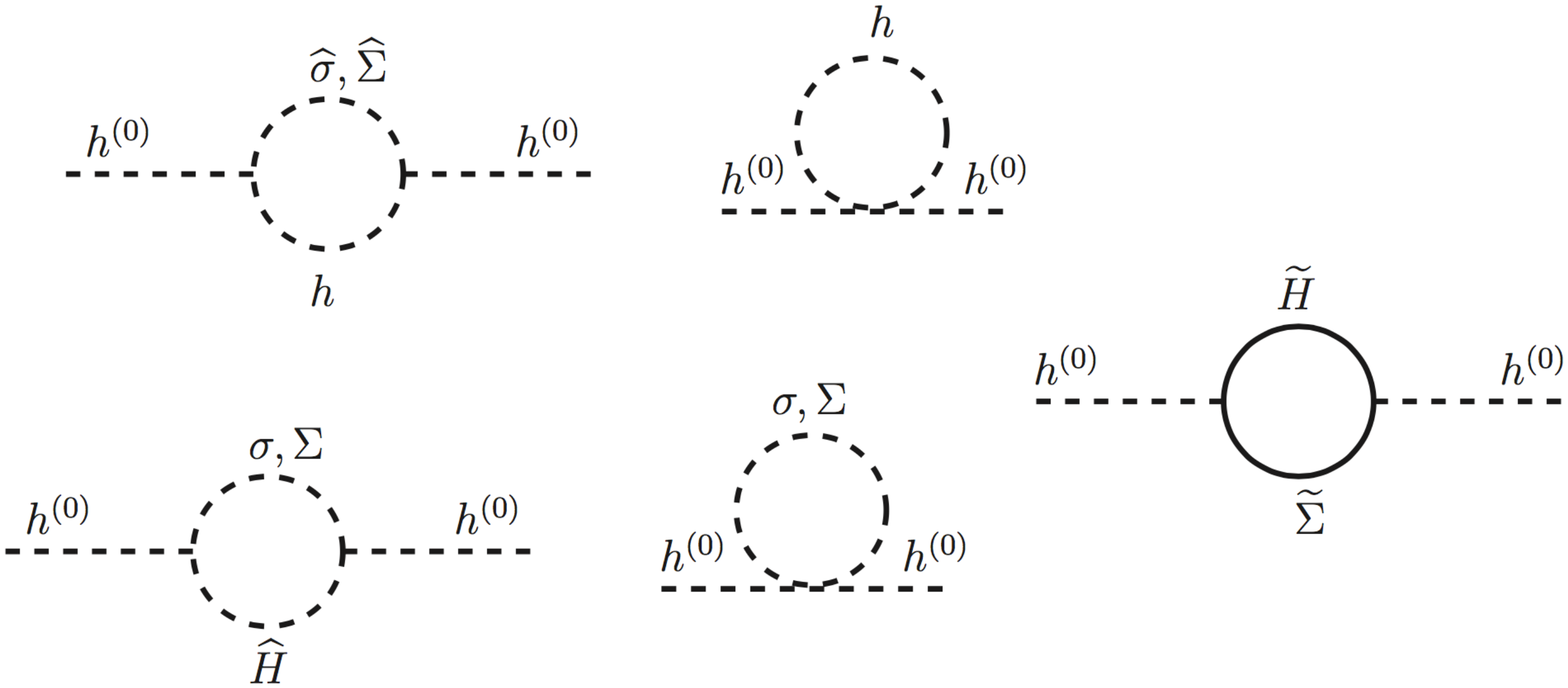}
\vspace{-2.5cm}\caption{\it Diagrams contributing to the mass term $\Delta_\lambda m^2 $.}
\label{diagrams}
\end{center}
\end{figure}

The EWSB is actually achievable in the present scenario. For our
choice of twists the tree-level squared mass $m^2$ is zero (see
Sec.~\ref{MSSM}). The radiative correction $\Delta_h m^2$ can be
split as
\be
\Delta_h m^2=\Delta_gm^2+\Delta_{\lambda}m^2\, ,
\ee
with $\Delta_gm^2$ provided in Eq.~\eqref{eq:Deltag}.  The
contribution $\Delta_{\lambda}m^2$ comes from the interactions
depending on the superpotential couplings $\lambda_{1,2}$. It is
produced via the diagrams in Fig.~\ref{diagrams} and results in
\begin{align}
  \Delta_{\lambda}m^2=&\frac{(\lambda_1-\lambda_2)^2+(\lambda_1+\lambda_2)^2}{2 (4 \pi)^4}~
  \widetilde\Omega(\omega)~,
\label{eq:Delta_l12}
\end{align}
with
\begin{align}
\widetilde \Omega(\omega)=&\Big\{2\zeta(3)-4 \Li_3\left( e^{2i\pi\omega}\right)
+4i\cot(2\pi\omega)\Li_ 4\left( e^{2i\pi\omega}\right)\nonumber\\
&+\Li_3\left( e^{4i\pi\omega}\right)-i\frac{2+3\cos(4\pi\omega)}{\sin(4\pi\omega)}\Li_4\left( e^{4i\pi\omega}\right)+h.c.\Big\} (1/R)^2~ .
\end{align}
%
%
%
%

\begin{figure}[h!]
\begin{center}
\hspace{-1.5cm} \includegraphics[width=5.8cm]{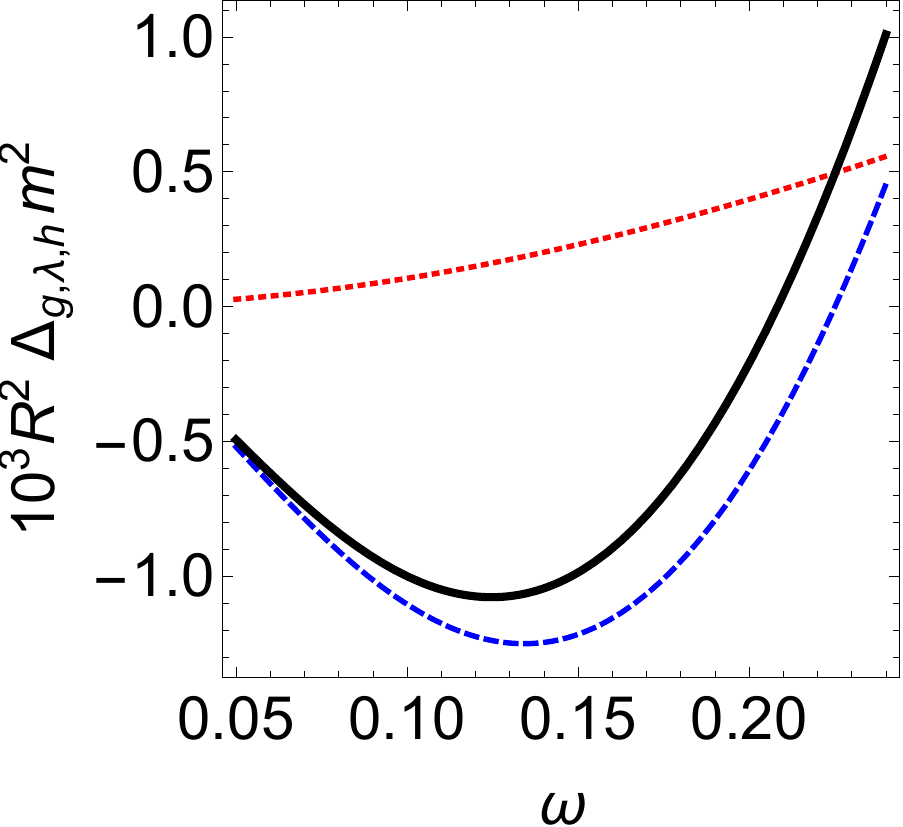}  \hspace{0.cm}
 \includegraphics[width=5.2cm]{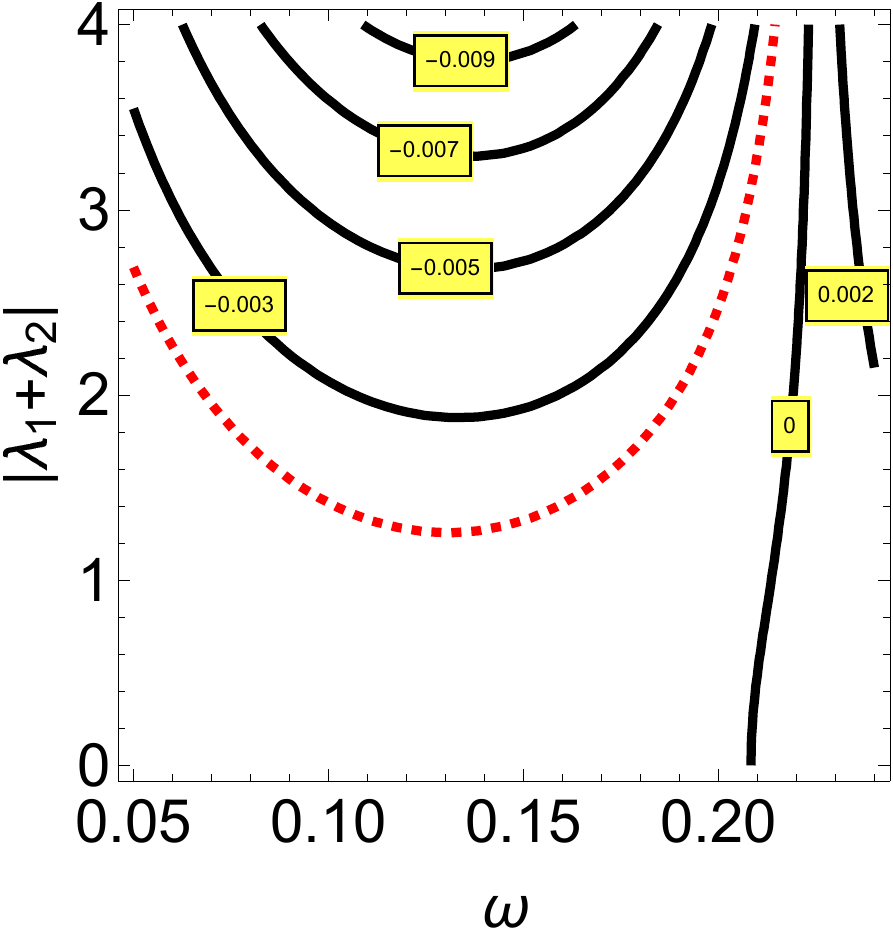}   \hspace{0.cm}
 \includegraphics[width=5.4cm]{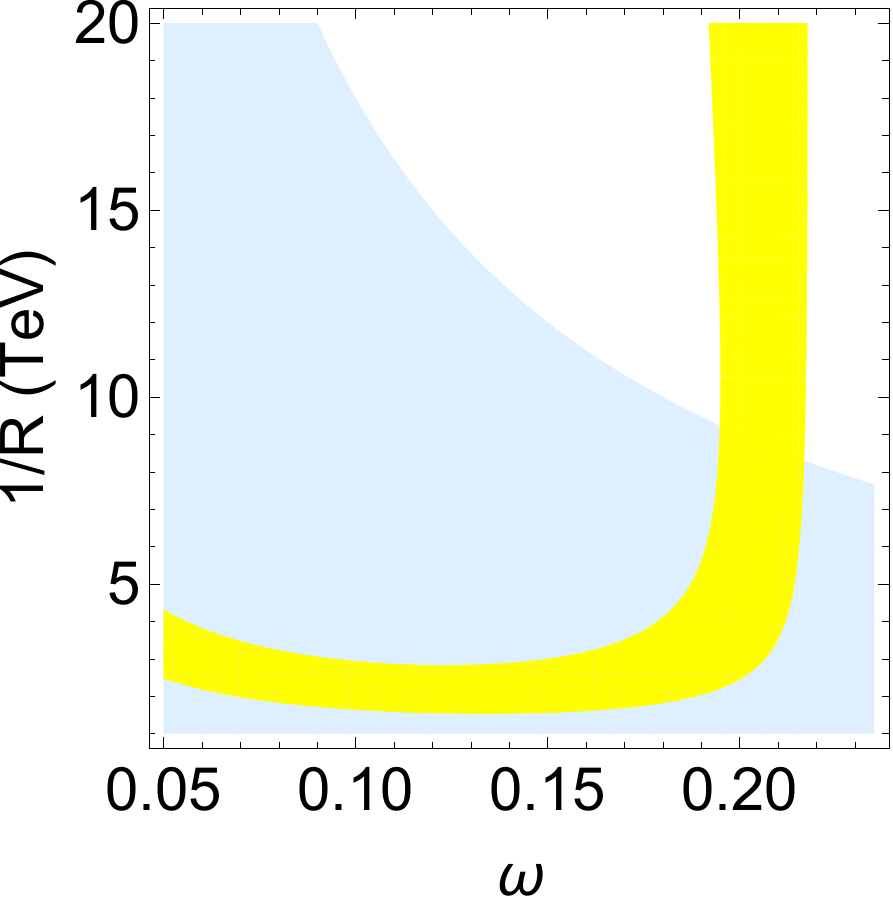}\hspace{-1cm}
\caption{\it Left panel: Plots of $\Delta_g m^2$ (red, dotted line),
  $\Delta_{\lambda}m^2$ (blue, dashed line) and their sum $\Delta_h
  m^2=\Delta_g m^2+\Delta_{\lambda}m^2$ (black, solid line) as
  functions of $\omega$ in units of $10^3R^2$ with $1/R=2\,$TeV and
  $\lambda_1+\lambda_2=0$.  Central panel: Contour plot of
  $R^2\Delta_h m^2$ in the $(\omega,|\lambda_1+\lambda_2|)$ plane for
  $1/R=2\,$TeV. The correct EWSB with the experimentally observed
  $125$ GeV Higgs mass happens along the dashed red line.  Right
  panel: The parameter space of the $(\omega,1/R)$ plane (yellow area)
  where the experimental EWSB with the correct Higgs mass is
  successfully achieved.  Blue area is as in Fig.~\ref{massesrad}. In all
  panels $|\lambda_1-\lambda_2|$ is fixed as in
  Fig.~\ref{lambdaminus}. }
 \label{fig:EWSB}
\label{Deltah}  
\end{center}
\end{figure}
Figure~\ref{Deltah} (left panel) displays the values of $\Delta_h m^2$
(solid line) and its contributions $\Delta_gm^2$ (dotted line) and
$\Delta_{\lambda}m^2$ (dashed line) in units of $10^3R^2$. The
plot highlights the illustrative case $1/R=2\,$TeV.  It assumes
$|\lambda_1-\lambda_2|$ fixed to reproduce the observed Higgs mass
(cf.~Fig.~\ref{lambdaminus}) and $\lambda_1+\lambda_2$ set to zero to
conservatively minimize the effect of $\Delta_{\lambda}m^2$ (see
Eq.~\eqref{eq:Delta_l12}). We see that $\Delta_g m^2$ is positive for
all values of $\omega$ whereas $\Delta_{\lambda}m^2$ can be
negative and sizable. In particular, at $\omega\lesssim 1/5$, $\Delta_h
m^2$ is negative and the EWSB is achieved. Of course, the larger the
value of $|\lambda_1+\lambda_2|$ the more easily the EWSB is
obtained. This is highlighted in the central panel of
Fig.~\ref{Deltah} presenting the contour lines of $R^2\Delta_h m^2$
in the $(\omega,|\lambda_1+\lambda_2|)$ plane with still
$1/R=2\,$TeV and $|\lambda_1-\lambda_2|$ fulfilling the Higgs mass
constraint. Along the (red) dotted line
the condition $\Delta_h m^2\approx -(88\,{\rm GeV})^2$ for the observed
EWSB is satisfied.
The finding is generalized to any value of $R$ in the right panel of
Fig.~\ref{fig:EWSB} where the yellow area highlights the region of
$(\omega,1/R)$ providing $\Delta_h m^2= -(88\,{\rm GeV})^2$ with
$|\lambda_1+\lambda_2|\le 2$ (the inner border corresponding to
$\lambda_1+\lambda_2=0$, the outer to $\lambda_1+\lambda_2=2$). Also
in this panel $|\lambda_1-\lambda_2|$ is adjusted to reproduce the
experimental Higgs mass.

We conclude that in this 5D embedding the presence of triplets
\begin{itemize}
\item
Enhances the tree-level Higgs quartic coupling in the effective theory
and thus makes it easy to accommodate the 125 GeV Higgs mass
constraint. This is essential for the naturalness of the model.
\item
Triggers the EWSB by providing a sizable negative contribution to the
squared-mass term of the SM-like Higgs and, in a somewhat wide
parameter region, generates the observed electroweak scale.

\end{itemize}

\subsection{Triplet trilinear term and Higgs-triplet quartic coupling}
Besides the squared mass of the triplet, there are other interactions
that are important for the phenomenology of the model. In this section
we focus on the triplet trilinear parameter and the Higgs-triplet
quartic coupling by which we can determine the VEV and masses of the
triplets (see Sec. \ref{sec:TripVEV}). 

The trilinear $A_\lambda$ term
\be
\mathcal L_4=\dots-A_\lambda h^\dagger \sigma h+h.c.
\label{eq:Alamb}
\ee
is generated by a loop of Higgsinos and gauginos. It can be evaluated
following the method employed for the stop mixing
$A_t$ in Ref.~\cite{Delgado:1998qr}. The result is given by
\be
A_\lambda=\frac{3(\lambda_1+\lambda_2)\alpha_2}{16\sqrt{2}\pi^2}C_2^h\left[  i \Li_2\left( e^{-2 i \pi\omega} \right)+h.c.\right]\frac{1}{R}\, ,
\ee
where $C_2^h=3/4$ is the quadratic Casimir of the Higgs doublet.
\begin{figure}[htb]
\begin{center}
\vspace{-5cm}
\includegraphics[width=12cm,angle=270]{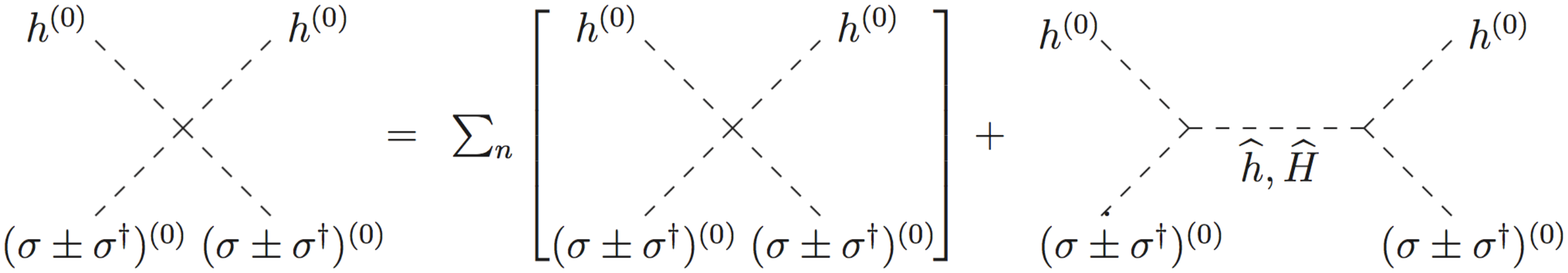}
\vspace{-4cm}
 \caption{\label{fig:matchCuart} \it High-energy diagrams leading to
   the low-energy quartic interactions between $h^{(0)}$ and
   $(\sigma\pm\sigma^\dagger)^{(0)}$, mediated by the propagation of
   $\widehat h$ or $\widehat H$, respectively. }
\end{center}
\end{figure}

In order to calculate the quartic coupling between the light Higgs
$h^{(0)}$ and the light triplet $\sigma^{(0)}$, we follow the
procedure used in Sec.~\ref{sec:quartic}. Specifically, we
determine the tree level matching of the Higgs-triplet interaction
between the high-energy theory described in Eq.~\eqref{eq:L4bis} and
the low-energy one where only the (tree level massless) zero modes
exist. As previously stated, we focus on the case
$\omega=q_R=q_H=q_\Sigma$.  The nontrivial contributions are those
corresponding to the vertices $|h^{(0)}|^2
|(\sigma+\sigma^\dagger)^{(0)}|^2$, mediated by the propagation of
$\widehat h$, and $|h^{(0)}|^2 |(\sigma-\sigma^\dagger)^{(0)}|^2$,
mediated by the propagation of $\widehat H$, depicted in
Fig.~\ref{fig:matchCuart}. They can be evaluated by means of the
identities
\begin{equation}
\lim_{p\to 0}\sum_n\frac{(\lambda_1\pm \lambda_2)^2}{16}\left[1+\frac{(q_R\mp q_H+n)^2}{p^2-(q_R\mp q_H+n)^2}\right]=\frac{(\lambda_1\pm\lambda_2)^2}{16}\delta_{q_R\mp q_H,0}\, ,
\label{sigma2Higgs2}
\end{equation}
and they lead to the following quartic interaction term:
\bea
\label{eq:quarticHsigma}
{\mathcal L}_{4D}&=&-\frac{(\lambda_1+\lambda_2)^2}{16} h^{(0)\dagger}[\sigma^{(0)}+\sigma^{(0)\dagger},\sigma^{(0)}+\sigma^{(0)\dagger}]_{+}\, h^{(0)}+\dots\nonumber\\
&=&-\frac{(\lambda_1+\lambda_2)^2}{8}h^{(0)\dagger}  h^{(0)}\sum_A[(t_R^A)^{(0)}]^2+\dots\, ,
\eea
with $(t_R^A)^{(0)}$ defined as
\be
\sigma^{(0)}=\frac{(t_R^A)^{(0)}+i (t_I^A)^{(0)}}{2}\tau^A
\label{eq:tRA}~.
\ee

\subsection{Triplet VEV and mass spectrum}
\label{sec:TripVEV}

The presence of the triplets has practically no effect on the mass
spectrum of the 5D MSSM-like states but $h^{(0)}$. In fact, all modes
of $\mathbb H_a$ except $h^{(0)}$ have large tree-level masses due to
the SS mechanism, and the remaining MSSM-like fields do not have
contact interaction with the triplets. The mass spectra shown in
Figs.~\ref{massesrad} and \ref{massestree} then hold correct also in
the present scenario (although only the (yellow) subregion highlighted
in the right panel of Fig.~\ref{fig:EWSB} is consistent with the
observed EWSB and 125 GeV $h^{(0)}$ mass for
$|\lambda_1+\lambda_2|\le2$). The relevant difference between the
spectra of the 5D MSSM and the triplet extension is then the masses of
the additional fields. 

\begin{figure}[htb]
\begin{center}
\vspace{-5cm}
\includegraphics[width=12cm,angle=270]{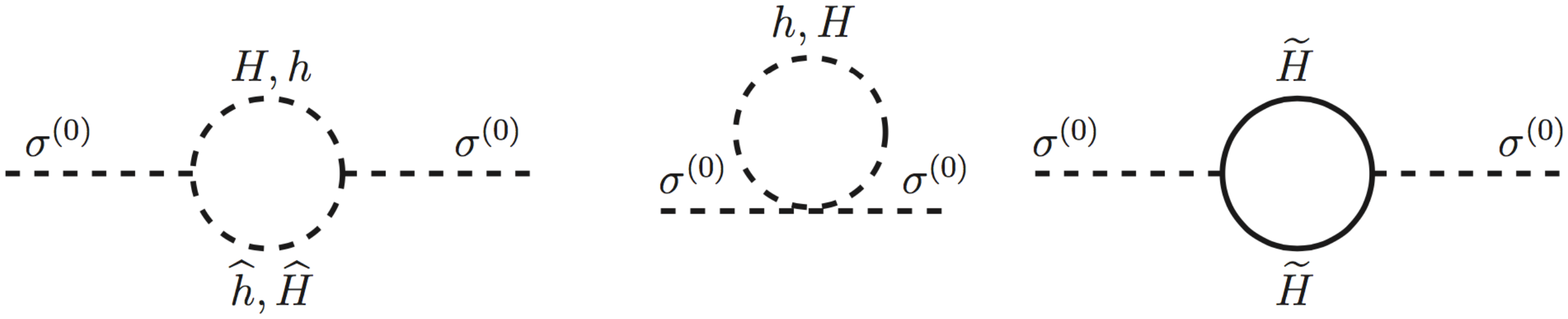}
\vspace{-4cm}\caption{\it Diagrams contributing to the mass term $\Delta m^2_\sigma|\sigma^{(0)}|^2$.}
\end{center}
\label{diagramsSigma}
\end{figure}
\begin{figure}[h!]
\begin{center}
  \includegraphics[width=7.5cm]{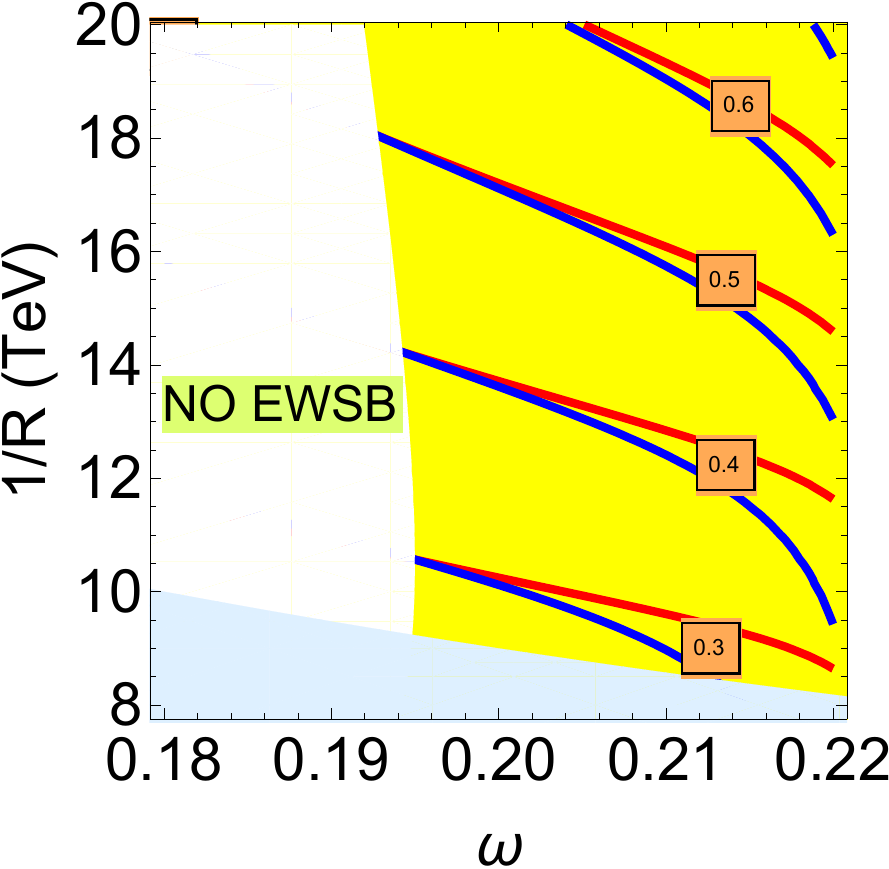}
  \hspace{0.7cm}
  \includegraphics[width=7.5cm]{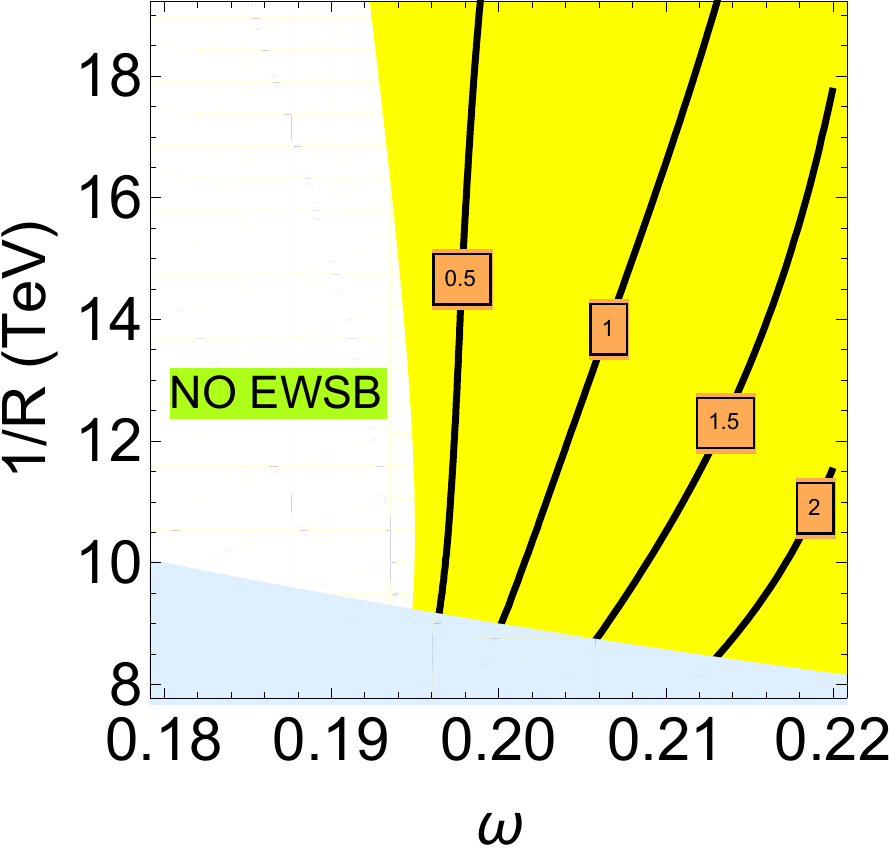}
\caption{ \it Left panel: Contour plot of the real $m_\sigma^R$ (blue
  curves), and imaginary $m_\sigma^I$ (red curves) triplet masses.
  Right panel: Contour plot of the VEV (in GeV) that the triplet
  acquires due to the loop-induced trilinear term.  In both panels it
  is assumed $|\lambda_1+\lambda_2|<2$ and the correct EWSB with a
  125 GeV Higgs mass is achieved inside the yellow region. Blue areas
  are as in Fig.~\ref{massesrad}. }
\label{tripletVEV}
\end{center}
\end{figure}

With respect to the SS twists, the fermionic components of the
triplets behave as the Higgsinos and they are hence degenerate in mass
with such fields. The scalar triplet $\sigma^{(0)}$ is instead insensitive to
the SS mechanism at tree level (for $q_R=q_H=q_\Sigma$), and only its real part 
$\tau_A(t_R^A)^{(0)}/2$ receives a mass by means of the
Higgs EWSB (cf.~Eq.~\eqref{eq:quarticHsigma}). However, this mass tends
to be subdominant with respect to the one coming from the one-loop
mass term $\Delta m^2_\sigma(\omega)|\sigma^{(0)}|^2$ that is produced
by the diagrams in Fig.~\ref{diagramsSigma}. This correction amounts
to
\begin{align}
  \Delta m^2_\sigma=&\frac{3(\lambda_1+\lambda_2)^2}{(4\pi)^4} \Omega^{+}_\sigma(\omega) +\frac{6(\lambda_1-\lambda_2)^2}{(4\pi)^4} \Omega^{-}_\sigma(\omega)\, ,
\end{align}
with
\begin{align}
\Omega_\sigma^-(\omega)=&\Big\{ -\Li_3\left(e^{2i\pi\omega}\right)+i\cot(2\pi\omega)\left[ -\Li_4\left(e^{2i\pi\omega}\right)+\Li_4\left(e^{4i\pi\omega}\right) \right]    
+h.c.\Big\}(1/R)^2\, , \nonumber\\
\Omega_\sigma^+(\omega)=&\Big\{ 2\zeta(3)-2
\Li_3\left(e^{2i\pi\omega}\right)+
\Li_3\left(e^{4i\pi\omega}\right)+2i\cot(2\pi\omega)\Li_4\left(e^{2i\pi\omega}\right)\nonumber\\
&-i\cot(4\pi\omega)\Li_4\left(e^{4i\pi\omega}\right)
+h.c.\Big\} (1/R)^2\,.
\end{align}
Therefore, after the EWSB, the squared masses of the zero modes of the
real and imaginary parts of the complex triplets are respectively
given by
\bea
(m_\sigma^R)^2&=&\frac{(\lambda_1+\lambda_2)^2}{4}v^2+\Delta m^2_\sigma~, \label{eq:Msigma} \\
(m_\sigma^I)^2&=&\Delta m^2_\sigma~,
\eea
The values of these masses as a function of $\omega$ and $1/R$ are
displayed in the left panel of Fig.~\ref{tripletVEV} where in the
yellow area $\lambda_1$ and $\lambda_2$ are fixed as usual to satisfy
the EWSB and Higgs mass constraints.

We now study the triplet VEV.  When the Higgs breaks the electroweak
symmetry, the trilinear interaction in Eq.~\eqref{eq:Alamb} induces a
tadpole term $\sim A_\lambda v^2 (t^3_R)^{(0)}$. This in turn induces
the VEV $\langle t_R^3\rangle\equiv \langle (t_R^3)^{(0)}\rangle$
which breaks custodial symmetry and affects the electroweak precision
observable $\rho$ as~\cite{Agashe:2014kda}
\be
\Delta\rho=\frac{4 \langle t^3_R\rangle^2}{v^2}\ ,
\ee
which, using the ($1\sigma$) bound $\Delta\rho\lesssim 6\times
10^{-4}$, provides the corresponding bound $ \langle
t^3_R\rangle\lesssim3$~GeV. Notice also the relation between the
observables $\Delta\rho$ and $T$ as given by $\Delta\rho=\alpha
T$~\footnote{A more complete description will require other
    observables correlated with $T$, as $S$ and
    $U$~\cite{Peskin:1991sw}. They will also be consistent with our
    model parameters as we will analyze at the end of
    Section~\ref{sec:low}.}.

The size of $\langle t_R^3\rangle$ is now obtained by considering the
scalar potential involving the tadpole term and the squared mass in
Eqs.~\eqref{eq:Alamb} and \eqref{eq:Msigma}. Its order of magnitude is
$\mathcal O(A_\lambda v^2/(m_\sigma^R)^2)$~\cite{Delgado:2012sm} and
its precise value is shown in the right panel of Fig.~\ref{tripletVEV}
where we plot $\langle t^3_R \rangle$ as a function of $\omega$ and
$1/R$ for $\lambda_1-\lambda_2$ fixed by the experimental Higgs mass
(cf.~Fig.~\ref{lambdaminus}). The finding is displayed only in the
(yellow) region where the observed EWSB can be achieved for
$|\lambda_1+\lambda_2|\le2$. The measurement of the $\rho$ parameter,
which imposes $\langle t^3_R\rangle\lesssim 2$
GeV~\cite{Delgado:2012sm, Agashe:2014kda}, provides no constraint on
the model besides in the corner with $\omega \gtrsim 0.21$ and
$1/R\lesssim10\,$TeV. In particular, along the left border of the
yellow area, which corresponds to $(\lambda_1+\lambda_2)=0$, no
trilinear term and thus no VEV of $(t_3^R)^{(0)}$ is generated. This
border, although fine-tuned, is technically natural as $\lambda_1$ and
$\lambda_2$ are supersymmetric parameters.

Finally, as a consistency check, we verify that the mixing between the
$h^{(0)}$ and $(t_3^R)^{(0)}$ is tiny. Otherwise our criteria
$\Delta_h m^2\simeq -(88\,\rm{GeV})^2$ and $m_{h}=125\,$GeV to
accomplish the observed EWSB and Higgs mass constraints would be
wrong. The mixing is sourced by the trilinear term in
Eq.~\eqref{eq:Alamb} after the EWSB. The mixing angle $\gamma$ can be
estimated as
\be
\tan(2\gamma)\sim \frac{A_\lambda v}{(m^R_\sigma)^2- m_h^2}\sim \frac{\langle t_3^R\rangle}{v}~,
\ee
where in the last step the Higgs squared mass has been neglected in
front of $(m^R_\sigma)^2$. The mixing is therefore fully negligible in
the whole area compatible with the electroweak precision observables
for which $\langle t_3^R \rangle\lesssim 3\,$GeV.

\begin{table}[htb]
\centering
\begin{tabular}{||c|c|c|c|c|c|c||}
\hline
$m_{\widetilde f_{1,2}}=m_{\widetilde\tau_R}=M_a=m_{\widetilde H}=m_{\widetilde \Sigma}$&$m_H=m_\Sigma$&$m_{\widetilde t_L}$&$m_{\widetilde t_R}$&$m_{\widetilde {\tau}_L}=m_{\widetilde {\nu}_\tau}$&$m_\sigma^R$&$m_\sigma^I$\\
\hline
 3000&6000&970&900&420&450&440\\
\hline
\end{tabular}
\caption{\it A sample of new-physics masses (in GeV) for $1/R=14$\,TeV
  and $\omega/R=3$ corresponding to $\omega\simeq 0.21$. The symbol
  $m_{\tilde f_{1,2}}$ represents the mass of all sfermions of the
  first and second generations. The radiative corrections are included
  only for the masses of the last five columns.}
\label{tabla1}
\end{table}
To conclude, we provide some explicit values for an illustrative
parameter scenario.  We consider the benchmark point with
$\omega=3/14$ and $1/R=14\,$TeV. In this case the EWSB and Higgs mass
constraints can be overcome with $\lambda_1\simeq 1.1$ and
$\lambda_2\simeq 0.0$. The triplet VEV, which turns out to be $\langle
t_3^R\rangle\simeq 1$\,GeV, is compatible with the above
$\rho$-parameter bound. The masses of the lightest modes corresponding
to new physics are quoted in Table~\ref{tabla1}. Some of them are
within the reach of the LHC although they can be elusive to the
standard searches as discussed in the next section.

\subsection{The low-energy phenomenology}
\label{sec:low}
Below the energy scale of the bulk fields with masses $\mathcal
O(\omega/R)$, the theory is described by the SM degrees of freedom
plus the scalar triplet $\sigma^{(0)}$, and the third-generation
squarks and slepton doublets.

In this setup the tau sneutrino ($\widetilde\nu_\tau$) is the LSP. The
$\widetilde\nu_\tau$ is not a good dark matter candidate as it would
provide the observed relic abundance for a (small) mass range that is
nevertheless ruled out by direct detection
experiments~\cite{Falk:1994es}. In the remaining mass region, its
relic density has to be somehow reduced. This is possible if the
sneutrinos do not reach thermal equilibrium before their freeze-out,
or an entropy injection occurs at late times (see
e.g.~\cite{Gelmini:2010zh, Nardini:2011hu }). Alternatively, decays
such as $\widetilde \nu_{\tau}\rightarrow \tau\bar{\tau}$ can provide
the desired dilution. These could in principle be generated by
operators like $LLE$ that introduce a small R-parity violation.

The collider phenomenology of the left-handed stau
($\widetilde\tau_L$) depends on its mass splitting with
$\widetilde\nu_\tau$. At tree level these fields are degenerate in
mass, and only QED one-loop corrections break the
degeneracy~\cite{Cirelli:2005uq}. For the part of the parameter space
that we are interested in, this splitting is at least $\mathcal
O(100\,{\rm MeV})$ and the lifetime of the stau is $\mathcal
O(0.1\,{\rm ns})$ or smaller, for which the ATLAS and CMS constraints
on disappearing tracks can be interpreted as ruling out
$m_{\widetilde{\tau}}<150$\,GeV~\cite{Aad:2013yna,CMS:2014gxa}. This
low-mass range is also ruled out by other constraints, as we now see.

Even though the gluino is not part of the low-energy theory, the most
robust constraint to the parameter space of the model is provided by
the gluino direct searches. From early 13 TeV data, ATLAS and CMS set
the bound $m_{\widetilde{g}}\gtrsim
1.8\,$TeV~\cite{ATLAS-CONF-2015-067,CMS-PAS-SUS-15-003}. Since the
whole spectrum mostly depends on just two parameters, $\omega$ and
$1/R$, and in particular $m_{\widetilde{g}}=\omega/R$, the gluino mass
bound constrains the low energy theory. The excluded region
corresponds to the blue areas in
Figs.~\ref{massesrad},~\ref{massestree},~\ref{plot:singlete},
\ref{lambdaminus}, \ref{fig:EWSB} and \ref{tripletVEV}. In particular,
as highlighted in Figs.~\ref{massesrad} and \ref{tripletVEV}, the
gluino bound forces the mass of the stops and sbottoms to be roughly
above $550$\,GeV and the scalar triplet, stau and tau sneutrino to be
heavier than around $250$\,GeV. Of course, by excluding heavier gluino
masses we will also be able to set stronger bounds on third-generation
squarks, the triplet and the stau doublet. However, the way in which
the tree-level generated gaugino masses scale with $1/R$ is different
from how the radiatively generated light states do. Even a $3$ TeV
bound on gluino masses will not be able to exclude stops at around
$1$\,TeV nor stau and triplet masses around $500$\,GeV (see
Table~\ref{tabla1}).  Hence it is worth studying also the
phenomenology of these particles.

As we are dealing with a heavy LSP with a mass typically above
$300$\,GeV, the LHC bounds from stop searches are very mild or even
absent~\cite{Aad:2015pfx}. In addition, considering usual bounds is a
conservative assumption; in this model the topology of the stop decays
is different from what is expected in MSSM-like scenarios. Because the
stop is lighter than all neutralinos and charginos, it decays to
off-shell states such that the final signature is a multibody decay
for which the current stop bounds can be very much
softened~\cite{Alves:2013wra}. Bounds on sbottoms are more severe than
those on stops (for LSP masses below $400$ GeV, ATLAS and CMS exclude
sbottom masses up to
$900$\,GeV~\cite{ATLAS-CONF-2015-066,CMS-PAS-SUS-16-001}) but they
suffer from the same softening mentioned above for stops. In this
sense, in the present model the phenomenology of the third-generation
squark is similar to the one analyzed in Ref.~\cite{Garcia:2015sfa}.

Direct detection of the scalar triplet is challenging. The
  triplet does not mix with the Higgs, is fermiophobic and gets a very
  small VEV; thus, production mechanisms such as gluon fusion or
  vector boson fusion are of no use. Multilepton searches can be
  employed, but these are able to constrain only the parameter space
  where the triplet is very light ($\lesssim 200\,$GeV) and acquires a
  VEV close to the $\rho$-parameter
  bound~\cite{Bandyopadhyay:2014vma}. Alternatively, by using
  Drell-Yan double production one can constrain fermiophobic scalars
  that do not acquire a sizable VEV and have no other way to be
  produced~\cite{Delgado:2016arn}. Because of kinematics, the
  Drell-Yan process gets weaker for larger triplet masses and to rule
  out masses above $250$\,GeV one would need a $100$ TeV collider in
  which the Drell-Yan production cross section is enhanced.

Finally, modifications to the loop-induced decay rates
$\Gamma(g\to\gamma\gamma)$ and $\Gamma(h\to Z\gamma)$ could be
generated by the new charged scalars of the triplet or the stau. These
can result in deviations of the Higgs signal strengths; however, they
will be very suppressed as very light masses ($\lesssim 200$ GeV) or
large couplings, $\mathcal{O}(1)$, are needed to produce a significant
enhancement in the Higgs decay rates. Similarly, the loop level
contributions to the S and T parameters are very small as the gluino
bound already forbids the new low-energy states to be below $300$ GeV,
where significant modifications could be generated. We have explicitly
calculated these using the results from Ref.~\cite{Khandker:2012zu}
and found no significant contributions in the parts of the $(\omega,
1/R)$ plane which are not already excluded by other measurements. In
particular, we find that $T^{\mathrm{1-loop}}<0.02$ and
$S^{\mathrm{1-loop}}<0.002$, well inside their experimental
bounds~\cite{Agashe:2014kda}.

\section{Conclusions}
\label{sec:concl}
In the present paper we have explored extra dimensions as a way to
minimize the fine-tuning triggered by the LHC constraints on minimal
supersymmetric extensions of the Standard Model. We have performed our
study focusing on five-dimensional supersymmetric embeddings, with the
fifth dimension compactified on an orbifold and $N=1$ supersymmetry
breaking of the Scherk-Schwarz type.

The Scherk-Schwarz paradigm for SUSY breaking has been extensively
explored in the literature and is able to provide interesting ways out
for some of the shortcomings of conventional scenarios of
softly broken supersymmetry. For instance, the $\mu/B_\mu$ problem is
avoided as a large Higgsino mass arises without any dimensional
parameter in the superpotential. The spectrum exhibits a pattern made
of compressed sectors, each one hierarchically separated in mass from
the others by multiples of $\omega/R$ or $1/R$ (with $\omega$ and $R$
being, respectively, the Scherk-Schwarz twist and the size of the extra
dimension). In this way the first- and second-generation sfermions are
naturally much heavier than the third-generation ones, in agreement
with flavor constraints. Moreover, due to the absence of large effects
in the renormalization-group evolution of the parameters, the
framework is also free of the `gluino-sucks' problem and sub-TeV
third-generation squarks are easily accommodated. The two main
drawbacks of the paradigm come when considering the electroweak
symmetry breaking and the experimentally measured Higgs mass, both not
achievable in minimal realizations of Scherk-Schwarz supersymmetry
breaking.

The present paper proves that these two problems are not generic
obstacles in Scherk-Schwarz scenarios. It shows that, for instance,
both issues can be solved in an extension with $Y=0$ $SU(2)_L$
triplets propagating in the bulk. It turns out that such triplets
both provide radiative corrections triggering the electroweak symmetry
breaking and enhance the tree level Higgs mass, so that the 125 GeV
mass is adjusted more naturally.

Because of the mass hierarchy between fields that propagate in the bulk and fields localized in the brane, most of the new-physics sector is decoupled
from electroweak-scale processes, in agreement with
experiments. However some superpartners, tightly linked to naturalness
and/or properties of the Scherk-Schwarz twists, have to be light and
populate the low-energy particle content of the theory, which
eventually consists of the Standard Model degrees of freedom plus a
scalar triplet, the third-generation of squarks and the doublet of sleptons. The
presence of the right-handed staus in the light spectrum, that we have avoided in the paper, is
really optional. Depending on the choice, the LSP can be the right-handed
stau or the tau sneutrino. The latter is preferable to avoid the
strong constraints on charged LSPs.

Since gluino bounds are robust and quite generic, the most stringent
constraint to the model comes from gluino searches. Nevertheless,
other experimental signals could be used to test it. In the short
term, searches for disappearing tracks or fermiophobic scalars are the
most promising for probing part of the parameter space. Searches for
the third generation of squarks are also important, but it is
challenging to apply their bounds to the present scenario where
squarks have multibody decays~\cite{Alves:2013wra}. We leave for the
future the reinterpretation of these bounds in terms of the parameter
space of the model.

\section*{Acknowledgments}
We thank ICTP-SAIFR for the kind hospitality during the first stages
of this article.  The work of A.D. is partly supported by the
National Science Foundation under Grant No. PHY-1520966. The work of G.N. is supported by the Swiss National Science Foundation (SNF) under Grant No. 200020-155935. The work of M.G.-P. and M.Q.~is partly supported by MINECO under Grant No.
CICYT-FEDER-FPA2014-55613-P, by the Severo Ochoa Excellence Program of
MINECO under Grant No. SO-2012-0234, and by Secretaria d'Universitats i
Recerca del Departament d'Economia i Coneixement de la Generalitat de
Catalunya under Grant No. 2014 SGR 1450. The work of M.Q. was also partly supported by CNPq PVE fellowship Project No. 405559/2013-5.  


\providecommand{\href}[2]{#2}\begingroup\raggedright\endgroup
\end{document}